\documentclass[11pt]{article}
\usepackage[a4paper,margin=1in]{geometry}
\usepackage{amsmath,amssymb,amsthm,mathtools,bm}
\usepackage{natbib}
\usepackage{xcolor}
\usepackage[hidelinks]{hyperref}
\usepackage{booktabs}
\usepackage{microtype}
\usepackage{enumitem}

\theoremstyle{definition}

\theoremstyle{remark}

\providecommand{\E}{\mathbb{E}}
\providecommand{\R}{\mathbb{R}}
\providecommand{\one}{\mathbf{1}}
\providecommand{\tr}{\operatorname{tr}}
\providecommand{\diag}{\operatorname{diag}}
\providecommand{\Var}{\operatorname{Var}}

\newcommand{\methodflaw}[1]{{\color{red!75!black}\textbf{[flagged:} #1\textbf{]}}}
\newcommand{\unverified}[1]{{\color{orange!80!black}[unverified: #1]}}
\providecommand{\unproven}[1]{{\color{red!70!black}[unproven: #1]}}
\usepackage{draftwatermark}
\SetWatermarkText{DRAFT}
\SetWatermarkScale{1.6}
\SetWatermarkColor[gray]{0.9}
\title{\textbf{Theorist Toolbox}\\[2pt]\large Tools for Agent Based LLM-assisted Economic Theory Research.
}
\author{Moran Koren\\ \small Ben-Gurion University of the Negev\\ \small \texttt{korenmor@bgu.ac.il}}
\date{June 2026}

\begin{document}
\maketitle

\begin{center}\fcolorbox{red}{red!4}{\parbox{0.93\linewidth}{\centering\color{red}\bfseries This is a draft. The formal model and the mathematical analysis in this paper are \emph{not} verified. They are included only to demonstrate the automated proof methods that are the subject of this paper, and should be read as such.}}\end{center}

\begin{abstract}
Empirical economists often start their projects with a toolbox. Shared packages, replication archives, and circulated guides shorten the time between and idea and a rough initial draft. Theorists, on the other-hand, largely start from a blank page. By 2026, large language models can a produce and check nontrivial mathematics. The can also hallucinate and write wrong claims very convincingly. The  current bottleneck on machine-assisted theory is no longer production but trust: a model will claim to prove a false theorem as readily as a true one. Building on recent attempts in mathematics, I present 3 methods for doing economic theory with a language model. These methods differ on how the work is verified: a single disciplined pass, an adversarial prover-verifier pair (Claude Opus~4.8 proposing, OpenAI Codex refuting), and a structured multi-agent project with a reviewer gate (inspired by the Google co-mathematician architecture). I demonstrate these protocols on one open worked example: designing a Groves/Pigouvian incentive mechanism for the Gans--Kominers eigengrade model of grade inflation.  None of the three runs produced a strict direct-revelation VCG/Clarke mechanism (as requested, perhaps due to the non-existence of such mechanism).  Three phenomena recur. First, convergent discovery: two runs derive the same effective-resistance externality kernel on opposite margins. Second, adversarial verification is load-bearing: the pair caught three of its own false claims and the gate rejected a sub-goal. Third, polish is not rigor: the most finished-looking output was the least verified. The methodological takeaway is that external verification, not model capability, is the design variable.
\end{abstract}

\section{Introduction}
\label{sec:intro}

A researcher beginning an applied project does not start from a blank page: there are packages,  and replication archives (e.g., \citep{gpswiv2020}, \citep{cunningham2021} etc.). The accumulated tooling is not incidental to applied economics; it is part of how the field transmits and disciplines its methods. In economic theory, the equivalent tool box is scarce. A theorist begins each proof from scratch and seldom, there exists a shared, reusable, semi-automatic infrastructure.

Large language models have the potential to reduce this asymmetry. Models are now capable enough at mathematics that they can carry a derivation and even sketch a proof. On the other hand, they are susceptible to hallucinations, and often take shortcuts. 
The difficulty therefore is not producing a seemingly clean, confident argument, but to guarantee its correctness. In economics, there is an additional challenge of making sure the assumptions made, and analysis don't render the mathematical formulation mute and deprived of any economic meaning.  

This paper proposes a method, meant to shorten the time between an idea and a rough initial draft. I document a \emph{verification-first protocol} for doing economic theory with language models, comprising three methods that differ on the manner in which result is verified mathematically. The economic significant of the results is still beyond reach. I evaluate all three on one mechanism-design problem. The evidence is a single worked example with one model pairing run by one operator, so the three phenomena I report below are proof of concepts that these effects \emph{can} occur. The three methods are: (i) a \emph{single careful pass}, in which one model writes a complete proof under a fixed discipline, with no code and no second opinion; (ii) an \emph{adversarial pair}, in which one model proposes a result and a second, independent model is instructed to break it, with the human triaging the findings and Monte Carlo checks running alongside; and (iii) a \emph{multi-agent project} that scaffolds the work the way one would run a research team, following the design of the AI co-mathematician of \citet{zheng2026}, with separate workstreams for literature, proving, and coding, a strict mode in which every gap is flagged, and a reviewer gate that must sign off before a result is counted as complete. The language model throughout is Claude Opus~4.8; the adversarial pair couples it with a second model (OpenAI Codex, gpt-5.5) acting as the hostile verifier.

The test problem is the worked example, not the contribution. It is a mechanism-design problem based on the grading model of \citet{ganskominers2026} which comprises an additive two-way model of latent performance and course difficulty that makes grading  manipulable. In the proposed problem, I ask whether a designer can deter manipulators using price mechanism rather than to estimate around the existing manipulation. The three methods, run independently on this problem, produced three distinct mechanisms, a student-side enrollment mechanism, an instructor-side Pigouvian transfer, and a money-free report-space suite with its own impossibility theorem. The reader interested only in grading will find usable results, but they are an unverified by-product. The contribution is the comparison and what it teaches about how the three methods behave.

The comparison surfaces three transferable phenomena, which I preview here and develop in Section~\ref{sec:findings}. First, \emph{convergent discovery}: two of the runs, which never saw each other's output, derived the same externality object, an effective-resistance term on the transcript-graph Laplacian, one pricing the extensive margin (which courses exist) and the other the intensive margin (how faithfully each is graded). When two runs arrive at the same non-obvious formula from different directions, the agreement raises confidence that the object is a feature of the problem and localizes the mathematics worth trusting. Second, \emph{adversarial verification is load-bearing}: the adversarial pair caught three of its own claims, including a budget-balance result that a single confident pass would have shipped, and the multi-agent project's reviewer gate rejected an entire sub-goal. A method that can reject its own output is worth substantially more than one that cannot. Third, \emph{polish is not rigor}: ordered by how finished they look, the single pass leads and the multi-agent project trails; yet the single pass carried no numerical check while the multi-agent output, the one covered in unproven flags, states what it does not know. The most finished-looking output was the least verified, and a reader of machine-assisted theory should read for the verification scaffolding rather than for the prose.

%The relevance to \emph{Management Science} is methodological. 

As researchers across economics, operations, and management increasingly use language models in the production of theory, the field needs protocols for doing so that are explicit about verification and reproducible across problems. The protocol here is not specific to grading or even to mechanism design; the three methods, and the three findings about how they behave, are intended to transfer to any problem in which one wants a machine to help with a proof without being permitted to hand-wave. To make the protocol usable and citable rather than merely described, the three methods are released as a small set of Claude Code skills.\footnote{The skills, the three example outputs, and companion materials are available at \url{https://github.com/morankor/theorist-toolbox}.} I am also explicit about the ceiling: none of the three methods proved a hard new theorem from nothing. What they produced together is a well-mapped problem with a verified core, a genuine impossibility result, a working money-free mechanism, and an explicit ledger of what remains open, which is the realistic deliverable from a machine at present and a useful one. I stress again, these results are unverified! But they cross the bar for an early draft of the proposed problem.

The remainder of the paper is organized as follows. Section~\ref{sec:problem} states the test problem and the grading model it builds on. Section~\ref{sec:methods} describes the three methods and the verification discipline each embodies. Sections~\ref{subsec:single}, \ref{subsec:adversarial}, and \ref{subsec:multiagent} present what the single pass, the adversarial pair, and the multi-agent project respectively produced on the problem. Section~\ref{sec:findings} states the three observed phenomena that constitute the paper's result. Section~\ref{sec:toolbox} closes with the released toolbox and guidance on composing the methods. Appendix~\ref{app:methods-output} catalogs the errors each method made and how each was caught, which is itself part of the evidence.

\section{Background and positioning}
\label{sec:background}

This paper sits at the intersection of three literatures that rarely meet: the use of large language models (LLMs) as instruments for mathematical research; the verification of machine-generated mathematics; and the mechanism-design theory of information elicitation. My contribution is a practitioner protocol for the first, disciplined by the lessons of the second, and demonstrated on a worked example drawn from the third. I therefore organize the background around these strands, and close by stating precisely how the present work differs from the formal-methods program that dominates the headlines.

\subsection{LLMs and agents for mathematics}
\label{subsec:llms-math}

The capability frontier has moved quickly enough that a position taken eighteen months ago is now stale. On the formal side, reinforcement-learning systems that operate inside an interactive theorem prover have reached competition-grade performance: DeepMind's AlphaProof, paired with AlphaGeometry~2, solved four of the six 2024 International Mathematical Olympiad problems at the silver-medal threshold, generating and checking candidate proofs in Lean and training on a corpus of roughly eighty million formalized propositions auto-translated from natural-language problems \citep{deepmind2024imo,alphaproof2025nature}. The defining feature of these systems is that correctness is \emph{machine-checked}: a proof is accepted only if the Lean kernel certifies it, so the search can be rewarded by a ground-truth verifier rather than by human or model judgment.

A parallel line treats the proof assistant as a target for translation rather than search. \emph{Autoformalization}---the conversion of informal statements and proofs into the language of Lean, Isabelle/HOL, or Coq---has become markedly more capable as LLMs trained on code and mathematics are combined with retrieval over libraries such as Lean's \texttt{mathlib} \citep{autoformalization2025survey,proofbridge2025}. Recent demonstrations are large in scale: substantial portions of Munkres' \emph{Topology} have been autoformalized in Isabelle/HOL and in the Megalodon system, the latter producing on the order of $130{,}000$ lines of formal mathematics in roughly two weeks \citep{munkres2026isabelle}. Tooling around Lean~4---interfaces to proof state, retrieval, and dedicated provers trained on synthetic data---has matured into an ecosystem rather than a single model \citep{autoformalization2025survey}.

Benchmarks have moved in step. Saturation of grade-school and competition datasets prompted Epoch~AI's \emph{FrontierMath}, a set of unpublished, expert-authored problems whose upper tier (Tier~4) is research-level and on which leading models long solved only a small fraction \citep{frontiermath2024,epoch_frontiermath}. The benchmark is itself a moving target: a substantial fraction of the original problems were later found to contain errors and were reissued in a corrected version \citep{epoch_frontiermath}, a fact worth recording because it illustrates the central difficulty of this paper---that even carefully curated mathematical artifacts harbor errors that surface only under adversarial scrutiny.

Closest to the present work is the shift from single-model proving toward \emph{agentic} research assistance. \citet{zheng2026} describe an ``AI co-mathematician,'' a stateful, asynchronous workbench in which AI agents support the full arc of a research project---ideation, literature search, computational exploration, theorem proving, and theory building---rather than discharging a single fixed goal; the system reports a $48\%$ score on FrontierMath Tier~4, a high-water mark among the systems compared there. One of the three production methods I study (Section~\ref{sec:methods}, ``multi-agent project'') adopts the architecture of \citet{zheng2026}: workstreams, a strict ``name every gap'' discipline, and a reviewer gate. I use it not as a benchmark contestant but as one organizational form among three, and I evaluate it on a single open theory question rather than on a problem set with known answers.

\subsection{Verifying machine-generated mathematics}
\label{subsec:verification}

The reason verification is the organizing concern of this paper, and not an afterthought, is that the dominant failure mode of LLM mathematics is not incoherence but \emph{fluent, plausible, and wrong}. A growing empirical literature documents the \emph{false-positive} problem: under answer-only grading, models receive credit for correct final answers reached through invalid intermediate reasoning, so reported accuracy overstates genuine competence \citep{falsepositives2025}. When an LLM is used as a judge, the problem compounds---judges reward solvers that fabricate algebraic facts to force a desired conclusion, in some cases assigning near-perfect scores to demonstrably false derivations \citep{qedbench2026}. Most directly relevant, \emph{sycophancy} benchmarks show that frontier models, when handed a false theorem, frequently produce a confident ``proof'' of it rather than flagging the error: on \textsc{BrokenMath}, the strongest model tested still returned sycophantic proofs of corrupted statements roughly $29\%$ of the time, and mitigation reduces but does not eliminate the behavior \citep{brokenmath2025}.

These findings motivate a design principle that recurs throughout this paper: a model's own confidence is uninformative about correctness, and verification must come from a source the prover cannot satisfy by agreement. In formal-methods systems that source is the kernel of a proof assistant. Outside formal methods---where economic theory currently lives, because its objects (equilibria, incentive constraints, welfare functionals) are not yet routinely formalized---the analogous discipline must be assembled from independent components: an adversary rewarded for finding counterexamples, numerical cross-checks of analytic claims, and an explicit ledger of unproven obligations. My protocol is an attempt to specify such a discipline and to measure what it buys, using the \texttt{methodflaw} and \texttt{unverified} conventions to mark, in the reproduced outputs themselves, exactly where each method's claims failed or remained unaudited.

\subsection{Information elicitation and peer prediction}
\label{subsec:peerprediction}

One of the three methods attempts, and its internal reviewer ultimately rejects, a sub-goal that would have elicited honest grades through a peer-prediction scheme; situating that failure requires the elicitation literature. The peer-prediction method of \citet{millerresnickzeckhauser2005} rewards an agent's report by how well it predicts a randomly chosen peer's report, and shows that under a common prior truthful reporting is a strict Bayes--Nash equilibrium even when no ground truth is ever observed---precisely the situation of grade honesty, where the ``true'' performance is latent. Two features of that result are load-bearing here. First, peer prediction relies on a genuine \emph{stochastic dependence} between an agent's private signal and the peer signal used to score it; when the scoring statistic is independent of the agent's own signal, the mechanism elicits the prior mean and a constant report weakly dominates honesty. Second, truth-telling is one equilibrium among many, and uninformative equilibria can pay at least as well---a fragility addressed by later work, including robust and small-population variants of the Bayesian truth serum \citep{witkowskiparkes2012}. The rejected sub-goal in Section~\ref{sec:methods} fails for exactly the first reason: its leave-one-out residual is constructed to be independent of the grader's own signal, so it cannot reward honesty. I flag this not as a defect of the method that produced it but as evidence that an adversarial reviewer gate, properly instituted, catches errors that the elicitation literature would predict.

\subsection{Mechanism design and the grading testbed}
\label{subsec:vcg-gk}

The worked example is a mechanism-design problem, so I place the relevant theory briefly. The Vickrey--Clarke--Groves (VCG) family makes truthful revelation a dominant strategy by charging each agent the externality her report imposes on the rest of society, aligning private and social optima \citep{vickrey1961,clarke1971,groves1973}. The construction carries a well-known cost: \citet{greenlaffont1979} establish that no mechanism can be simultaneously efficient, dominant-strategy incentive-compatible, and budget-balanced, so a VCG mechanism generically runs a deficit. This impossibility is not an obstacle to be engineered away in the worked example; it is the organizing tension, and the three methods discharge it in three different currencies (deficit, a Bayes--Nash prior, and a market-power tax).

The testbed itself is the grading model of \citet{ganskominers2026}. They write latent performance of student~$i$ in course~$c$ as $P_{ic}=a_i-d_c+\varepsilon_{ic}$, with $a_i$ a student-ability effect and $d_c$ a composite course effect (difficulty, grading standard, instructor strictness, cohort), and show that a scalar grade cannot separate the two: cross-course ability rankings can reverse at a grade cutoff whenever course effects are large enough (their impossibility, Thm.~4.5). Their constructive remedy treats the transcript as a two-way fixed-effects problem on the bipartite enrollment graph---a signed incidence matrix $B$, Laplacian $L=B^{\top}B$, and recovered centered ability $L^{+}B^{\top}p$ on a connected graph---and they show that row-mean, affinity-spectral, and graph-Laplacian methods recover the \emph{same} object, so the ``eigengrade'' is a representation of fixed-effect adjustment rather than a new source of identification. Two features make this an interesting  VCG testbed. First, the precision of the recovered abilities is a global spectral quantity (their Prop.~6.15, $\mathbb{E}\lVert\hat{\alpha}-(a-\bar a\mathbf{1})\rVert^{2}\le\sigma^{2}\,\mathrm{tr}(L^{+})\le \sigma^{2}(n+m-1)/\lambda_2(L)$), so every enrollment and every grading choice imposes a \emph{computable} informational externality on everyone else. Second, the adjustment delivers strategic \emph{neutrality} but not \emph{efficiency}: exact difficulty adjustment makes student course choice neutral and shuts the competitive-enrollment channel of instructor grade inflation, but it neither rewards informative enrollment nor guarantees that the raw inputs are honest. Those two gaps are the openings the worked example exploits.

\subsection{What this paper is, and is not}
\label{subsec:positioning}

The contribution is a \emph{practitioner protocol} for doing economic theory with LLMs, together with a controlled comparison of three ways of running it, in a domain---economic theory---where machine-checked formalization is not yet available. This distinguishes the work from the autoformalization and proof-search programs of Section~\ref{subsec:llms-math} in three ways. First, the objects of interest are economic (mechanisms, equilibria, welfare and incentive functionals), and they are verified not by a proof-assistant kernel but by a composite of adversarial counterexample search, numerical Monte Carlo, and an explicit obligation ledger. Second, the unit of study is the \emph{production process}---solo pass, adversarial pair, multi-agent team---rather than a model or a benchmark score; the comparison, not the mechanism, is the result. Third, the deliverable I argue for is not a certified theorem but a \emph{disciplined research frontier}: a verified core, working mechanisms, an honest negative result, and a named list of what remains open. I do not claim formal verification, I do not introduce a new prover or benchmark, and I treat every reproduced model output as a lead to be checked rather than a result to be trusted---a stance the verification literature of Section~\ref{subsec:verification} shows to be necessary rather than cautious.

\section{Three methods as a protocol}
\label{sec:methods}

The methodological claim of this paper is narrow and operational: when one delegates economic theory to a large language model, the organizing decision is not which model to use, nor how to prompt it, but \emph{how the work is verified}. Everything else---notation, exposition, even which result to chase---is downstream of that choice. This section presents three ways of doing LLM-assisted theory as three points on a single axis, the verification axis, ordered from least to most external scrutiny. The first method performs no verification beyond the disciplined self-checking of a single pass; the second introduces an adversarial second model whose job is to refute; the third wraps the work in a multi-agent project with a standing reviewer gate. The same underlying model---Claude Opus~4.8---drives all three. What differs is the apparatus around it.

I stress at the outset that these are not three ad hoc prompts. Each is a reusable Claude Code \emph{skill}: a versioned bundle of instructions, scripts, and conventions that the model loads when invoked, so that the same protocol can be re-run on a different problem without reconstruction.\footnote{The skills are \texttt{math-proof}, \texttt{codex-math}, and the \texttt{co-math-init}/\texttt{co-math-status} pair, distributed in the Theorist Toolbox repository (\url{https://github.com/morankor/theorist-toolbox}).} The protocol is the artifact; the worked example of Section~\ref{subsec:vcg-gk} is merely what the protocol was pointed at. For each method I give the setup, the roles (who proposes and who checks), the inputs and outputs, the verification mechanism, and the conditions under which it is the right tool.

\subsection{Method 1: a single disciplined pass}
\label{sec:method1}

The first method is a single pass of one model, governed by a writing discipline and checked by no one but itself. It is implemented as the \texttt{math-proof} skill, which loads a standard for proof writing rather than a problem-solving strategy. The standard is exacting in a specific way: no gaps between steps, every derivative signed and the sign justified, every definition bridged to its later use, intermediate algebra shown rather than asserted, the general case proved rather than inferred from small examples, and---most relevant to verification---an explicit instruction to distinguish what is proved from what is conjectured and to flag any load-bearing claim that rests on a plausible but unestablished step. The skill forbids the rhetorical moves (``it is easy to see,'' ``clearly,'' ``the other case is similar'') that let a proof skip the steps most likely to hide an error.

\paragraph{Roles, inputs, outputs.} There is one role. The model both proposes the mathematics and is its only critic, the critic being the same forward pass operating under the discipline above. The input is a claim or a proof sketch together with the model primitives; the output is a single self-contained proof document. In the worked example this was one long derivation of a student-side enrollment mechanism, with no code and no second opinion (Appendix~\ref{app:m1}).

\paragraph{Verification mechanism.} The only verification is internal consistency under the writing standard. The skill's value is that its prohibitions raise the cost of self-deception: signing every term forces the model to confront the steps it would otherwise wave through, and the explicit ``proved versus conjectured'' instruction yields a record of self-flagged limitations. This is real and worth having---in the worked example the single pass correctly refused to claim budget balance and correctly flagged that its individual-rationality argument depended on the choice of target. But self-flagging has a structural ceiling. A model cannot flag an error it does not perceive, and the discipline that makes the prose airtight is the same discipline that makes an unsound result \emph{look} airtight. No internal standard substitutes for an adversary or a numerical check; there was neither here. The cleanest, most finished-looking output in this study came from this method, and it was the least verified of the three (Section~\ref{sec:findings}).

\paragraph{When to reach for it.} Use the single pass when the result is one self-contained derivation, when the cost of a latent error is bounded (an exploratory lemma, a step you intend to verify downstream), or as the first stage feeding either of the heavier methods. It is fast and produces publication-grade prose. It should not be the last word on any load-bearing claim, because nothing in it can catch an error the model is confident about.

\subsection{Method 2: an adversarial pair}
\label{sec:method2}

The second method adds a second model whose institutional role is to \emph{refute}. It is implemented as the \texttt{codex-math} skill, which wires OpenAI Codex (\texttt{gpt-5.5}) into the workflow as an adversarial co-processor running non-interactively through \texttt{codex\,exec}. Codex offers three modes---verify a proof step by step, write a proof of a stated result, and explore whether a conjecture is true and under what conditions---of which the skill's own documentation marks exploration the most valuable, because it is the mode that returns counterexamples and boundary conditions rather than a verdict on prose already written.

\paragraph{Roles, inputs, outputs.} There are three roles. Claude Opus~4.8 proposes; Codex, prompted at high reasoning effort, attacks; and the human author triages. The triage role is not optional decoration but the load-bearing element, for a reason the skill states plainly and I adopt as the operating rule of this method: Codex is \emph{brilliant and unreliable}, and every output is a lead, not a verdict. It produces a substantial fraction of false positives---flagging standard conventions as errors, demanding unnecessary generality, over-reading degenerate cases---alongside its genuine catches, and the two are distinguishable only by inspection. A real catch comes with a concrete counterexample or points to a specific step that does not follow; a false positive is a convention mismatch or a demand for regularity conditions that are already standard. The skill's instruction is therefore explicit: do not accept or reject a Codex finding without checking the specific algebra yourself. The inputs are a proposed result and its proof; the outputs are the surviving result plus a transcript of attempted refutations and, in the worked example, two NumPy Monte Carlo audits that put the analytical claims against simulated data.

\paragraph{Verification mechanism.} Verification here is adversarial: a claim survives only if an independent model, trying to break it, fails, and a human has confirmed that the failure is genuine rather than a missed catch. This is the first method in the sequence that can overturn the proposer. In the worked example it overturned the proposer three times, killing three claims the author initially believed. Codex produced an explicit $K_{2,2}$ counterexample that demoted an ex-post incentive-neutrality claim to neutrality in expectation only; it proved budget balance \emph{impossible} via an $m$-fold mixed derivative the proposer had not computed; and it downgraded a claim that a leave-one-out construction was needed for first-moment exactness once it showed the standard estimator already had that property. These corrections are not incidental to the method; they are the method working as designed (Section~\ref{sec:findings}, Appendix~\ref{app:m2}).

\paragraph{When to reach for it.} Reach for the adversarial pair on a single hard result whose correctness matters and where a counterexample, if one exists, would be decisive: an existence or uniqueness claim, a proposed mechanism's incentive property, a conjecture you cannot settle by direct attempt. The skill's own guidance is to escalate to Codex precisely when a direct proof attempt has failed or when a routine self-check has flagged an unproved load-bearing claim, and to keep routine symbolic computation out of it. The pair is heavier than a single pass and narrower than a full project: it verifies a slice deeply rather than a program broadly.

\subsection{Method 3: a multi-agent project}
\label{sec:method3}

The third method scales verification from a single result to a research program by giving the work a project structure with a standing reviewer gate. It is implemented as a skill pair: \texttt{co-math-init} scaffolds a new project and \texttt{co-math-status} renders its state. The architecture follows the AI co-mathematician of \citet{zheng2026}, adapted to a single human principal coordinating specialized agents. Initialization creates a directory with a living \texttt{paper.tex}, a \texttt{goals.md} that must be approved before work begins, a \texttt{decisions.md} ledger, and a \texttt{workstreams/} registry; the status skill replaces that paper's workstream-branching figures with an ASCII rendering of goals, running and blocked workstreams, pending reviews, and open obligations.

\paragraph{Roles, inputs, outputs.} The project defines several agent roles---project coordinator, literature reviewer, prover, coder, and a paper reviewer---each a distinct invocation with a distinct remit, with the human as principal. Work is decomposed into sub-goals dispatched to workstreams, and Codex is called in as a sub-step for the hard adversarial checks, so Method~2 sits inside Method~3 as a component. The inputs are a research question and an approved goal set; the outputs are a compiling paper, per-workstream reports, a decisions ledger, and a set of reviewer verdicts written as approval files.

\paragraph{Verification mechanism.} Verification operates at two levels. First, \emph{strict mode}, the default at initialization, requires the prover to mark every gap with an \verb|\unproven{}| flag and forbids hand-waving for any step, so the document carries an explicit ledger of what it has not established---collected, in the worked example, into an appendix of surviving flags. Second, a \emph{reviewer gate}: completion requires approval, and each sub-goal receives a verdict of approve, approve-with-flags, or reject, recorded to disk. The gate has teeth in both directions. It forced an over-claimed existence result back to a conditional statement once the reviewer found a divergent counterexample, and it \emph{rejected} an entire sub-goal---a peer-prediction route---on the ground that the proposed estimator elicited the prior mean rather than honest reporting, so a constant report beat truthfulness; the rejected sub-goal survives in the artifact as an explicit negative result rather than being quietly dropped (Section~\ref{sec:findings}, Appendix~\ref{app:m3}). This is the distinguishing trait of the method: it names what it does not know, and the naming survives the project running to completion.

\paragraph{When to reach for it.} Reach for the multi-agent project when the question is a program rather than a single theorem---when there are several sub-results that interact, a literature to map, numerics to maintain alongside proofs, and a need for an audit trail of what was decided and what remains open. It is the heaviest of the three and the slowest to set up, and its characteristic failure mode is breadth without depth, an interrupted run that has opened more workstreams than it has closed. Its compensating virtue is that the strict-mode flags and the reviewer gate make that failure mode \emph{visible} rather than hidden, which is exactly what the lighter methods cannot guarantee.

\subsection{The axis}
\label{sec:axis}

Read in order, the three methods trace a single dimension. Method~1 verifies by internal discipline alone; Method~2 adds an external adversary and a human triage; Method~3 adds a project structure, a standing flag discipline, and a reviewer gate, with the adversarial pair embedded as a component. Cost and latency rise along the axis, and so does the strength of the guarantee: from ``the prose is internally consistent and the model flagged what it noticed'' to ``an independent model tried to break this and a human confirmed it failed'' to ``a program of results was decomposed, each gap is named, and each sub-goal carries a recorded verdict.'' The protocol is the recognition that this is a choice to be made deliberately---matching the verification apparatus to the cost of being wrong---rather than a property of whichever prompt one happened to type. Section~\ref{subsec:vcg-gk} runs all three on one problem so that the axis can be read off concrete outputs, and Section~\ref{sec:findings} reports what the comparison revealed.

\section{The test problem (worked example)}
\label{sec:problem}

A methodological claim about a research process is only as credible as the problem it is tested on. A protocol that succeeds on a toy exercise tells us little; one that succeeds on a problem an expert would find genuinely hard, and that is partly open even to the expert, tells us a great deal. The worked example in this paper is therefore not an illustration chosen for tractability. It is a stress test, deliberately selected because it is concrete, mathematically nontrivial, and unresolved in the source literature. This section sets it up. I first present the grading model of \citet{ganskominers2026} and the identification problem at its core; I then state the mechanism-design question the three methods of Sections~5 and~6 attack; and I close by explaining why this particular question is a good discriminator between a protocol that produces rigor and one that produces only its appearance.

\subsection{The Gans--Kominers grading problem}

A university transcript is asked to compress many heterogeneous classroom experiences into a portable signal of student quality. \citet{ganskominers2026} argue that the central obstacle is not grade inflation per se but an identification problem: a raw grade is a one-dimensional report generated in a setting with at least two relevant dimensions, a student component and a course component, and these are not separately identified from the grade alone.

Formally, they study an additive latent-performance benchmark. When student $i$ takes course $c$, latent cardinal performance is
\begin{equation}
P_{ic} = a_i - d_c + \varepsilon_{ic}, \qquad \mathbb{E}[\varepsilon_{ic}] = 0, \quad \operatorname{Var}(\varepsilon_{ic}) = \sigma^2,
\label{eq:gk-model}
\end{equation}
with the idiosyncratic shocks independent across student--course pairs and independent of $(a_i, d_c)$. Here $a_i$ is a scalar \emph{student effect}, read as the one-dimensional ability component relevant to the measurement model; $d_c$ is a scalar \emph{course effect}, a composite course intercept that may capture intrinsic difficulty, grading standards, assessment design, instructor strictness, or cohort strength. A grading rule maps performance through common thresholds into a finite ordered set of grade labels. Employers and graduate schools nevertheless use the resulting records to compare students across courses.

The difficulty is that a scalar grade cannot separate $a_i$ from $d_c$. A universally calibrated rule applies the same performance thresholds in every course, so the ability required to clear a given threshold varies with the course effect; a high grade can reflect high ability, a favorable course, a lenient standard, or some combination. \citet{ganskominers2026} make this precise through a grade-boundary reversal: a lower-ability student in a sufficiently favorable course can clear a grade cutoff that a higher-ability student in a sufficiently unfavorable course misses, and any such reversal violates the requirement that a higher grade imply at least as high ability. In the additive case, a boundary overlap across a cutoff $t_r$ is the event $a_{i_L} - d_{c_L} \ge t_r > a_{i_H} - d_{c_H}$ for some pair with $a_{i_H} > a_{i_L}$, and ability sufficiency holds on a realized curriculum exactly when no such overlap exists. The impossibility result is thus a curriculum-span statement: grade-only comparability fails precisely when course heterogeneity is large enough to swamp ability differences at a grade boundary. The disease is identification; grade inflation is one symptom.

\paragraph{The eigengrade fix.} The constructive part of \citet{ganskominers2026} recasts transcript adjustment as a two-way fixed-effects measurement problem on the bipartite student--course enrollment graph $\mathcal{G} = (S \cup C, E)$, with $|S| = n$ students and $|C| = m$ courses. Stack the observations into a signed incidence matrix $B$ whose row for edge $(i,c)$ carries $+1$ in the student column $i$ and $-1$ in the course column; every such row is orthogonal to the all-ones vector. Form the graph Laplacian $L = B^\top B$ and compute
\begin{equation}
\hat{x} = L^{+} B^\top p, \qquad x = (a, d),
\label{eq:eigengrade}
\end{equation}
where $L^{+}$ is the Moore--Penrose pseudoinverse and $p$ stacks the observed scores. On a connected graph the centered student effects $a - \bar a \mathbf{1}$ are recovered up to a single additive constant; this is the object \citet{ganskominers2026} call the \emph{eigengrade}. They emphasize that the spectral language is not a new source of identification. Row-mean, affinity-spectral, and graph-Laplacian constructions recover the same object, and eigengrades are a representation and implementation layer for fixed-effect adjustment, not a separate identifying assumption. What does the identifying work is connectedness of the enrollment graph, which lets cross-course comparisons propagate through shared students and courses rather than requiring every student to take every course.

\paragraph{Each report is an externality.} The precision of the recovered abilities is a global property of the graph spectrum. \citet{ganskominers2026} bound the mean-squared recovery error of the centered student effects $\tilde\alpha$ by
\begin{equation}
\mathbb{E}\bigl\lVert \tilde\alpha - (a - \bar a \mathbf{1}) \bigr\rVert^2 \;\le\; \sigma^2 \operatorname{tr}(L^{+}) \;\le\; \sigma^2 \,\frac{n + m - 1}{\lambda_2(L)},
\label{eq:precision-bound}
\end{equation}
where $\lambda_2(L)$ is the algebraic connectivity of $\mathcal{G}$. This bound is the conceptual seed of every mechanism considered below. Because the right-hand side depends on the whole Laplacian spectrum, every enrollment decision and every grade report changes the recovery precision of \emph{everyone's} measured ability, not just the reporting agent's. The estimation problem is shared, and each agent's action is an externality on it.

\paragraph{Neutrality, not efficiency.} Finally, \citet{ganskominers2026} study strategic response, and their results are statements of \emph{neutrality} rather than \emph{efficiency}. Under exact difficulty adjustment the public report becomes a function of $a_i$ rather than of $a_i - d_c$, so there is no direct signaling return to taking an easier course: the spread across courses in expected market valuation, their transcript mechanism's strategic susceptibility $\Omega(\mathcal{R})$, falls to zero. On the instructor side, exact adjustment makes the report-mediated demand pass-through, the semi-elasticity $\eta_{\mathcal{R}}$, vanish, which eliminates the competitive enrollment channel of grade inflation. In their reduced-form instructor payoff
\begin{equation}
U_c = \alpha \bar G_c + \beta \log s_c - \tfrac{\xi}{2}(\bar G_c - \bar G_0)^2,
\label{eq:instructor-payoff}
\end{equation}
the competitive-inflation term that exact adjustment removes is $\beta \, \eta_{\mathcal{R}} (1 - 1/m)/\xi$. The key word is removes: exact adjustment switches off the \emph{wrong} incentive but supplies no \emph{right} one, and the whole construction presumes that the raw scores fed into \eqref{eq:eigengrade} are reported honestly. Two gaps remain open: the informational externality of \eqref{eq:precision-bound} is left un-internalized, and the honesty of the inputs is assumed rather than enforced.

\paragraph{The running example.} \citet{ganskominers2026} carry a single illustrative example throughout: five students and eight courses, with each student taking exactly four courses on a connected enrollment graph. The scalar additive benchmark \eqref{eq:gk-model} is the special case in which the within-family ability variation is switched off. On this example the eigengrade markedly outperforms naive aggregation. The correlation between a recovered score and true ability is $0.812$ for letter-grade GPA and $0.837$ for the raw-score average, but $0.996$ for the graph-Laplacian eigengrade.\footnote{These three correlations are the headline numerical comparison in the \citet{ganskominers2026} running example; the multi-agent method of Section~6 reproduces them, which I note where relevant.} The example matters here because it is the natural validation harness: any mechanism built on top of the eigengrade can be checked against the same five students and eight courses, and a method that claims a result but never runs this check has left an obvious verification gap.

\subsection{The mechanism-design question}

The two open gaps suggest a mechanism. Course difficulty $d_c$ is a manipulable nuisance on both margins of the model. Students manipulate it on the \emph{extensive} margin by choosing low-difficulty courses; instructors manipulate it on the \emph{intensive} margin by setting lenient or compressed grades. In each case the manipulation is an externality on the shared estimation problem of \eqref{eq:precision-bound}, and in each case the eigengrade construction merely neutralizes the private return without aligning the agent with the planner.

There is a standard tool for exactly this configuration. The Vickrey--Clarke--Groves mechanism \citep{vickrey1961,clarke1971,groves1973} implements the socially efficient decision in dominant strategies by charging each agent the externality her report imposes on everyone else; once an agent internalizes the full social consequence of her action, her private optimum coincides with the planner's and truth-telling becomes dominant. The Gans--Kominers environment is an unusually clean candidate for this logic, because \eqref{eq:precision-bound} makes the social objective a computable function of agents' actions: each enrollment and each grading choice moves $\operatorname{tr}(L^{+})$, the total transcript imprecision, in closed form. The instinct, then, is not to estimate \emph{around} the nuisance, as the eigengrade does, but to \emph{price} it with a VCG transfer.

This is the question handed to the three methods. Concretely: can the eigengrade estimator of \citet{ganskominers2026} be turned into an incentive-compatible mechanism that internalizes the informational externality \eqref{eq:precision-bound}, on either the student margin (which courses are taken) or the instructor margin (how faithfully grades are reported)? A scoping note is in order before the methods are turned loose, because it bears on how their outputs should be read. None of the three runs produced a strict direct-revelation VCG/Clarke mechanism: two produced Groves/Pigouvian transfers over agents' \emph{actions} (enrollments and grades) rather than over reported types, and the reading of the eigengrade itself as ``a Groves mechanism'' is an analogy with no money, allocation rule, or quasilinear payoff identity behind it, as the adversarial pass explicitly established. I therefore reserve ``VCG'' for the general mechanism-design idea and for the with-money Groves benchmark of the third method, and describe the produced mechanisms as Groves/Pigouvian. The question carries a genuine twist on textbook VCG. Universities have no monetary numéraire to hand instructors per grade, and \citet{groves1973}-style transfers, together with the budget impossibility of efficient dominant-strategy mechanisms, presuppose transferable utility. Whether the externality can instead be priced in a money-free report-space numéraire is precisely the kind of question on which a principled method and a merely fluent one diverge.

\subsection{Why this is a good stress test}

Three features make the question a discriminating test of the verification-first protocol rather than a showcase for it.

First, it is \emph{concrete}. The model \eqref{eq:gk-model}, the estimator \eqref{eq:eigengrade}, and the precision bound \eqref{eq:precision-bound} are fully specified, and the running example supplies a numerical harness with a known answer ($0.996$ versus $0.812$). A proposed mechanism can therefore be checked both analytically and computationally, so a method that omits the computation has visibly skipped a step that was available to it.

Second, it is \emph{genuinely hard}. Building a VCG-style mechanism here requires differentiating $\operatorname{tr}(K L^{+} K^\top)$ through a pseudoinverse, handling the constant in the kernel of $L$, and confronting the classical tension among efficiency, dominant-strategy truthfulness, and budget balance. These are the kinds of steps where a confident single pass can ship a cancellation mistaken for budget balance, or an in-expectation property mistaken for an ex-post one, errors that are subtle precisely because they are plausible. The problem is thus rich enough to expose the failure modes the protocol is designed to catch.

Third, and most importantly, it is \emph{partly open}. \citet{ganskominers2026} stop at neutrality and explicitly leave the efficiency question and the honesty-of-inputs question unaddressed; the no-money variant has no textbook answer at all. Because there is no published solution to anchor on, a method cannot succeed by retrieval, and a method that only \emph{looks} rigorous cannot be rescued by a known target. An open problem is the setting in which the gap between polish and rigor is widest, and therefore the setting in which a protocol built to expose that gap earns its keep. The three phenomena reported in Section~6 (convergent discovery of a single externality kernel, the load-bearing role of adversarial verification, and the gap between polish and rigor) all depend on having posed a problem where these distinctions have real stakes. This one does.

\section{What each method produced}
\label{sec:outputs}

I ran the three protocols of Section~\ref{sec:methods} on a single, deliberately
open prompt: design a VCG-style mechanism for the Gans--Kominers grading model.
The prompt fixed the target paper and the model class but left the \emph{side} of
the market (students or instructors), the choice of numéraire, and the precise
efficiency notion to the method. The three runs therefore did not converge on a
restatement of one mechanism; each settled on a different object. This section
reports what each produced and states the central results at the level of clean
claims, faithful to the artifact each run left behind. Full proofs, and the
errors that the verification layers caught, are deferred to
Appendix~\ref{app:methods-output}; here I flag the load-bearing failures in passing
because they matter for the comparison in Section~\ref{sec:findings}.

Throughout, I keep the source paper's primitives: latent performance
$P_{ic}=a_i-d_c+\varepsilon_{ic}$ with $\E[\varepsilon_{ic}]=0$,
$\Var(\varepsilon_{ic})=\sigma^2$; students $S=\{1,\dots,n\}$, courses
$C=\{1,\dots,m\}$; signed incidence vectors $b_{ic}$ (a $+1$ in student
coordinate $i$, a $-1$ in course coordinate $c$, so $b_{ic}^\top(a,d)=a_i-d_c$),
the enrollment-graph Laplacian $L=\sum b_{ic}b_{ic}^\top$, its pseudoinverse
$L^+$, and the eigengrade precision bound
$\E\|\tilde\alpha-(a-\bar a\one)\|^2\le\sigma^2\tr(L^+)\le\sigma^2(n+m-1)/\lambda_2(L)$
\citep{ganskominers2026}. The recurring object across all three runs is the
effective-resistance form $\|\Pi L^+ b_{ic}\|^2$, where $\Pi$ projects onto the
estimation target; I return to its independent rediscovery in
Section~\ref{sec:findings}.

\subsection{Single pass: a student-side enrollment Groves mechanism}
\label{subsec:single}

The single disciplined pass (Method~1, the \texttt{math-proof} skill: one Claude
Opus~4.8 pass, no code, no second opinion) produced a \emph{student-side}
mechanism that prices the enrollment externality. Its starting observation is
that eigengrades buy \emph{neutrality}, not \emph{efficiency}: under exact
difficulty adjustment the strategic susceptibility is $\Omega(\R)=0$ and course
choice collapses to the purely pedagogical benchmark, but the precision bound
shows that every student's eigengrade variance is a function of the entire
enrollment graph through $L^+$. When student $i$ enrolls in course $c$ she adds
edge $(i,c)$, perturbs $L$, and changes the estimation error of \emph{everyone's}
transcript. This is a pure informational externality that the neutral rule
internalizes nowhere.

The mechanism closes the gap with a Groves transfer denominated in a currency
held \emph{outside} the transcript (enrollment priority points, tuition credits),
which must not be written into grades. Fixing a target contrast matrix $K$ whose
rows lie in $\one^\perp$ (the leading case is the shared course-contrast target,
$K$ the centered course effects), the run defines the \emph{informational cost}
of a graph as the total best-linear-unbiased-estimator variance of the target,
\[
  \Psi(E)\;=\;\tr\!\big(K\,L(E)^{+}K^\top\big)
  \quad(\text{connected }E),\qquad \Psi(E)=+\infty\ \text{otherwise,}
\]
which for the centered-student target reduces to $\tr(L^+)$ and reproduces the
paper's bound. The institution is a public agent with valuation
$V_0(E)=-\lambda\sigma^2\Psi(E)$; social surplus is
$W(E)=\sum_{j}v_j(E)+V_0(E)$, where $v_{ic}\ge0$ is student $i$'s private
pedagogical value of course $c$. Students report $\hat v$, the planner picks
$E^\star(\hat v)\in\arg\max_E\widehat W(E)$, and each student receives a Groves
transfer with the Clarke pivot $h_i=\mathcal W(S\setminus i)$.

The run states and proves six results. A structural lemma establishes that $\Psi$
is edge-monotone (adding an observation weakly lowers cost), participation-monotone
for the course-contrast target (adding a student's edges weakly lowers cost), and
convex in fractional enrollment intensities, with marginal informativeness
\[
  \frac{\partial\Psi}{\partial z_{ic}} \;=\; -\,\big\|K L^{+} b_{ic}\big\|^2 \;\le\;0 .
\]
On this base it proves: truthful reporting of pedagogical values is weakly
dominant (Groves); the mechanism implements the first-best curriculum,
internalizing the enrollment externality; under the Clarke pivot each student's
utility equals her social marginal contribution and is nonnegative
(individual rationality) \emph{for the course-contrast target}; the transfer is
exactly Pigouvian, with marginal subsidy equal to the true marginal social
informativeness $\lambda\sigma^2\|KL^{+}b_{ic}\|^2$ and \emph{no} $-d_c$ arbitrage
term; and the outcome strictly welfare-dominates the neutral eigengrade curriculum
whenever pedagogy and informativeness disagree. The continuous relaxation is an
A-optimal experimental-design program, hence convex and solvable to global
optimality. Two limitations are stated in the artifact itself: the mechanism runs
a Green--Laffont budget deficit (efficiency, dominant strategies, and budget
balance are jointly impossible for this public-good objective), and the
clean individual-rationality proof relies on participation-monotonicity, which
holds for the fixed-dimensional course-contrast target but
\methodflaw{fails for the student-ability target, where an added student
introduces her own variance term and the monotonicity argument breaks.}

The decisive feature of this output for the comparison to follow is that it is
the most polished of the three and carries \emph{no numerical verification at
all}: every claim is an analytic derivation, checked by nothing but the pass that
produced it.

\subsection{Adversarial pair: an instructor-side Pigouvian transfer}
\label{subsec:adversarial}

The adversarial pair (Method~2, the \texttt{codex-math} skill: Claude proposes,
OpenAI Codex (gpt-5.5) at high effort attempts to break each claim, I triage, and
a NumPy Monte~Carlo checks the survivors) attacked the \emph{instructor} side. It
delivers a genuine Groves/Pigouvian transfer that implements professional grading
in \emph{strictly dominant strategies} while keeping raw grades, together with a
sharpening of the student-side incentive result.

The instructor environment is the paper's \S8 game: course $c$ chooses mean grade
$\bar G_c$, logit enrollment shares $s_c(\bar G)=e^{\eta\bar G_c}/\sum_k e^{\eta\bar G_k}$
($\eta=\gamma\rho_\R\ge0$ the report-mediated pass-through, $\eta=0$ under exact
eigengrades), and intrinsic payoff
$U_c=\alpha\bar G_c+\beta\log s_c-\tfrac{\xi}{2}(\bar G_c-\bar G_0)^2$. The unique
symmetric equilibrium inflates the professional standard by the
competitive-enrollment term $\beta\eta(1-1/m)/\xi$, pure business-stealing since
shares sum to one. The mechanism keeps raw grades and pays instructor $c$ the
transfer
\[
  t_c(\bar G) \;=\; -\beta\log s_c(\bar G)\;+\;h_c(\bar G_{-c}),
\]
with $h_c$ a function of the other instructors' grades only. The main theorem is
that the $\beta\log s_c$ terms cancel exactly, so the residual objective is
strictly concave with maximizer $\bar G_c^\star=\bar G_0+\alpha/\xi$ independent
of $\bar G_{-c}$, $\eta$, and $\beta$. The professional profile is therefore the
unique strictly dominant-strategy equilibrium even when $\eta>0$: the
competitive-inflation term is removed by \emph{pricing} the enrollment externality
rather than by changing the signal. The implemented grading coincides with the
$\eta=0$ eigengrade equilibrium, so pricing the externality and removing it from
the signal are two routes to the same outcome. The run also recasts eigengrades
themselves as the \emph{informational} half of VCG (a Groves-form correction that
cancels the course nuisance using cross-student data, made literal by a
leave-one-out ``pivotal eigengrade'' that is self-influence-free), while flagging
that this is an analogy, not a formal equivalence, since there is no money or
allocation rule in the estimator.

What makes this run informative for the methodology is the set of claims the
adversary \emph{killed}. The author's transcript records three corrections, each
forced by Codex or the Monte~Carlo rather than conceded voluntarily:
\begin{enumerate}[leftmargin=1.6em,label=\textup{(\alph*)}]
  \item \methodflaw{An \emph{ex-post} incentive-neutrality claim for the pivotal
  eigengrade was disproved by a small $K_{2,2}$-style counterexample: realized
  other-student shocks make any single transcript sample-dependent, so only
  \emph{expected} neutrality survives, downgrading ``exact in every sample'' to
  ``exact in expectation.''}
  \item \methodflaw{A budget-balance claim was proved \emph{impossible}: the
  $m$-fold mixed derivative
  $\partial_{1\cdots m}\big(\beta\sum_c\log s_c\big)=\beta(-1)^m m!\,\eta^m\prod_c s_c=\beta(-1)^m m!\,\eta^m/m^m\neq0$
  (the closed form evaluated at the symmetric professional profile $s_c=1/m$)
  for $m\ge2,\beta,\eta>0$, so no smooth dominant-strategy transfer balances the
  budget; AGV expected-externality transfers restore balance only in Bayes--Nash.}
  \item \methodflaw{A claim that the leave-one-out construction was \emph{needed}
  for first-moment exactness was downgraded: the standard eigengrade is already
  first-moment neutral (confirmed by Monte~Carlo, $z=0.27$); the leave-out
  estimator's genuine value is self-influence-freeness, the literal Clarke-pivot
  property, not first-moment correctness.}
\end{enumerate}
These are not blemishes on the output but the point of the method: the verified
core (the dominant-strategy Pigouvian transfer) survived precisely because three
adjacent, plausible-looking claims did not.

\subsection{Multi-agent project: a money-free report-space suite}
\label{subsec:multiagent}

The multi-agent project (Method~3, the \texttt{co-math-init}/\texttt{co-math-status}
skills, with the workstream-and-reviewer architecture of \citet{zheng2026}) ran in
strict mode with a reviewer gate and produced not a single mechanism but a
\emph{suite} under one binding constraint chosen at scaffolding: no monetary
transfers; everything must be implemented in report space. The organizing
object is the weighted-Laplacian informativeness welfare. With instructors
injecting within-course reporting noise of variance $\tau_c^2$, the institution
runs weighted GLS with edge precisions $w_{ic}=(\sigma^2+\tau_c^2)^{-1}$,
$L_W=B^\top W B$, and target projector $\Pi=\diag(M_n,0_m)$; welfare is
$W(\tau)=-\tr(\Pi L_W^+\Pi)$. The benchmark workstream (SG0) derives the
externality kernel
\[
  \frac{\partial W}{\partial w_{ic}} \;=\; \big\|\Pi L_W^+ b_{ic}\big\|^2\ge0,
  \qquad
  \frac{\partial W}{\partial\tau_c}
  \;=\;-\,\frac{2\tau_c}{(\sigma^2+\tau_c^2)^2}\!\!\sum_{i:(i,c)\in E}\!\!\|\Pi L_W^+ b_{ic}\|^2\le0,
\]
and states the with-money Groves benchmark (dominant-strategy, efficient) as the
frontier the money-free mechanisms are measured against. This kernel is the same
effective-resistance form the single pass derived on the extensive margin; I
return to the coincidence in Section~\ref{sec:findings}.

The suite's positive results are four. On the \emph{instructor} side (SG1), a
fixed-supply enrollment-share rule with softmax penalty
$s_c(\tau)=e^{-\eta\psi_c(\tau_c)}/\sum_k e^{-\eta\psi_k(\tau_k)}$ reproduces the
money pivot up to a \emph{simplex tax} $1-1/m$: the private marginal value of
fidelity is the $(1-1/m)$ fraction of its social value, so the institution must
inflate intensity by $m/(m-1)$ to match money. This $1-1/m$ is exactly the
zero-sum factor in the source paper's competitive-inflation term
$\beta\eta_\R(1-1/m)/\xi$ \citep{ganskominers2026}. A follow-up (SG1-a) proves the
share game is an \emph{exact concave potential game} with potential
\[
  \Phi(\tau)\;=\;\sum_{c}\big[u_c(\tau_c)-\beta\eta\,\psi_c(\tau_c)\big]
              \;-\;\beta\log\!\sum_{k}e^{-\eta\psi_k(\tau_k)},
\]
whose gradient reproduces each instructor's private marginal incentive and whose
Hessian is negative semidefinite under ``$u_c$ concave, $\psi_c$ convex,''
delivering global best response and (under coercivity) a unique equilibrium for
arbitrary heterogeneous instructors. On the \emph{student} side (SG2), a
fixed-budget Eisenberg--Gale pseudo-market prices the extensive-margin
connectivity externality with the \emph{same} kernel $\|\Pi L^+ b_{ic}\|^2$ and
incurs the parallel tax $1-1/K$ across $K$ students. The project then proves its
own \emph{impossibility} (SG4): in the symmetric fixed-supply share class with an
intensity cap $\bar\eta$ derived from a share-feasibility/IR floor, full-fidelity
grading $\tau=0$ is implementable if and only if
$u'(0)\le\beta\bar\eta\,\psi'(0)(1-1/m)$; above this threshold the direct
grade-preference dominates and \emph{no} mechanism in the class implements full
fidelity. This is the money-free analogue of the paper's grade-boundary wall, with
the bounded fixed-supply numéraire playing the role of the bounded grade scale.
Finally, a composition result (SG5) proves existence of a pure-strategy
equilibrium of the two-sided game via Kakutani--Fan--Glicksberg and signs the
cross-margin interaction (own-course complementarity, generic-but-not-universal
cross-course substitution). The numerics workstream reproduces the source paper's
running example exactly: the graph-Laplacian eigengrade correlates with true
ability at $0.996$, against $0.812$ for raw GPA and $0.837$ for the raw-score
average.

Two outcomes of the reviewer gate are essential to the comparison. First, the
SG1-a \emph{existence} claim was overclaimed and the reviewer forced it back to a
coercivity/compact-box condition: \methodflaw{under only ``$u_c$ concave,
$\psi_c$ convex'' the share game can have \emph{no} equilibrium (an explicit
$m=2$, $u_c=2t$, $\psi_c=0.1t$ counterexample sends $\Phi$ to $+\infty$ along the
diagonal), so existence is recovered only on a compact action box or under a
coercivity hypothesis.} The same review corrected the conjectured potential, which
had wrongly contained the welfare term $W$; the true potential contains none.
Second, the peer-prediction sub-goal (SG3, truthful performance revelation) was
\methodflaw{\emph{rejected} by the reviewer gate: the leave-one-out residual is
statistically independent of the grader's own signal, so it elicits the prior
mean rather than the private signal, and a constant report equal to the prior mean
strictly out-scores ``honest'' reporting ($-1.130$ versus $-1.348$ in the verified
$5\times4$ instance). Only the correct independence proposition was retained; the
propriety and likelihood-alignment theorems were downgraded to open conjectures
and the section marked partial.} The remaining open obligations---integral
rounding of the pseudo-market, the exact asymmetric pivot, closed-form own-course
complementarity, composite uniqueness and a welfare bound---are collected,
unhidden, in the project's own appendix of \texttt{\textbackslash unproven} flags.
This is the raggedest of the three outputs on its surface, and, as
Section~\ref{sec:findings} argues, the most trustworthy.

\section{Findings: the three-method comparison}
\label{sec:findings}

The worked example of Sections~\ref{subsec:single}--\ref{subsec:multiagent} is, deliberately, an instrument. Three productions of the protocol --- a single disciplined pass (P1), an adversarial prover--verifier pair (P2), and a structured multi-agent project (P3) --- were aimed at one open research prompt: turn the Gans--Kominers eigengrade estimator into an incentive-compatible mechanism. Each produced a usable theory artifact, and those artifacts are reported elsewhere in the paper. The contribution of this section is not any one of them. It is what the \emph{comparison} reveals about doing economic theory with large language models. I organize that around three observed phenomena: that agreement between runs raises confidence as replication does; that adversarial verification was load-bearing rather than decorative; and that the visible polish of an output carried no information about its actual verification. These are demonstrations from one worked example, not measured rates --- a caveat I make in full in Section~\ref{sec:limitations} and keep in view throughout. I close with a short ``production-function'' reading of the three organizational forms.

Throughout, I tag results by their origin: \textbf{[P1]} for the single pass, \textbf{[P2]} for the adversarial pair, \textbf{[P3]} for the multi-agent project. The point of the tags is that the observations below are claims about \emph{which method produced what}, and they are checkable against the artifacts.

\subsection{Phenomenon 1 (observed): convergent discovery as cheap replication}
\label{sec:finding-convergent}

The single most reassuring fact in the exercise is that two of the three runs, working separately and never seeing each other's output, derived the same central object. One caveat must be attached to the word ``independently'' before any inferential weight is placed on it: both P1 and P3 are driven by the same base model, Claude Opus~4.8, so their agreement on the effective-resistance kernel shares whatever biases that model carries and is not two fully independent draws. The genuinely cross-model case in the exercise is the adversarial pair P2, which couples Claude with a second engine (OpenAI Codex). The convergence reported here is therefore same-model, different-framing convergence, and I read it accordingly. P1 and P3 were given different framings, used different instruments (money versus report-space), and disciplined different margins of behavior --- and both arrived at an effective-resistance externality kernel on the transcript Laplacian:
\begin{equation}
\underbrace{\frac{\partial}{\partial w_e}\operatorname{tr}\!\bigl(K L_W^{+} K^{\top}\bigr)}_{\text{P1: }d(\text{cost})}
\;=\; -\,\bigl\|K L_W^{+} b_e\bigr\|^{2},
\qquad
\underbrace{\frac{\partial}{\partial w_e}\Bigl(-\operatorname{tr}\!\bigl(\Pi L_W^{+}\Pi\bigr)\Bigr)}_{\text{P3: }d(\text{welfare})=d(-\text{cost})}
\;=\; +\,\bigl\|\Pi L_W^{+} b_e\bigr\|^{2},
\label{eq:kernel-convergent}
\end{equation}
where $L_W=\sum_e w_e\,b_e b_e^{\top}$ is the (weighted) bipartite Laplacian, $b_e$ the signed incidence vector of edge $e=(i,c)$, and $\Pi$ (resp.\ $K$) the centering projector (resp.\ contrast) defining the recovered ability target. The two sides now differentiate objects that differ by a sign \emph{by construction} --- P1 differentiates an estimation cost $\operatorname{tr}(K L_W^{+}K^{\top})$ and P3 the welfare $W=-\operatorname{tr}(\Pi L_W^{+}\Pi)$ --- so the opposite signs on the right are bookkeeping; the shared substance is the squared projected pseudoinverse $\lVert\cdot L_W^{+}b_e\rVert^{2}$ itself. Up to the choice of target operator these are the \emph{same} squared projected pseudoinverse applied to $b_e$ --- the effective resistance of edge $e$ in the transcript graph. P1 reaches it on the \emph{extensive} margin, differentiating the informational cost $\Psi(z)=\operatorname{tr}(K L(z)^{+}K^{\top})$ with respect to enrollment intensities $z_e$ (which courses a student takes). P3 reaches it on the \emph{intensive} margin, differentiating the informativeness welfare $W(\tau)=-\operatorname{tr}(\Pi L_W^{+}\Pi)$ with respect to the precision $w_{ic}=(\sigma^2+\tau_c^2)^{-1}$ an instructor puts on an edge (how faithfully she grades).

Both derivations turn on the same structural fact, and identifying it is what makes the agreement non-coincidental rather than two independent guesses that happened to match. Every incidence row satisfies $b_e^{\top}\mathbf 1 = 1-1 = 0$, so $\mathbf 1\in\ker L_W$ universally; in the differential of the pseudoinverse the correction term $(\partial L_W)(I-L_W L_W^{+})$ carries a factor $(b_e^{\top}\mathbf 1)\mathbf 1^{\top}$ that therefore vanishes. What survives is the clean rank-one update $\partial L_W^{+} = -L_W^{+}b_e b_e^{\top}L_W^{+}$, from which both expressions in \eqref{eq:kernel-convergent} follow. The two runs, without shared state, located the same load-bearing structural identity --- orthogonality of incidence rows to the constant vector --- and built the same object on it.

For an economist the inferential move is a familiar one: replication under different specifications. A large language model is a stochastic system, and a single run that derives a non-obvious closed-form expression is weak evidence that the expression is correct rather than a fluent confabulation. Two runs, on different sides of the problem, with different instruments and no shared state, landing on the identical formula raises confidence that the mathematics is real. I do not attempt to quantify how unlikely such a coincidence is: with one worked example and a shared base model that is not estimable, and the convergence is same-model rather than cross-model. What the agreement does is more modest and still useful --- it raises the posterior that the object is a feature of the problem rather than of the run, and it localizes the part of the derivation worth trusting. That is the epistemic role replication plays in empirical economics: it does not prove the result, it concentrates belief.

This yields a concrete protocol prescription, and it is cheap. To trust a model's non-obvious derivation, do not run it once; run it more than once, ideally under deliberately different framings, with different tooling, and --- the strongest version --- on a different base model, and check whether the runs converge on the same object. Convergence is near-free corroboration; divergence localizes where to look for the bug. P1 and P3 were not designed to cross-check each other --- they were aimed at different mechanisms --- which is why their agreement is informative rather than engineered; the residual caveat is that they share a base model, so the cleanest test, re-deriving the kernel on a second model, is left to future work.

The convergence also did genuine mathematical work that neither run did alone. P1 priced the extensive margin and P3 priced the intensive margin, each within its own framing, and neither presented its kernel as a special case of a margin-independent law. Seeing the two side by side reveals that \emph{one} kernel governs \emph{both} margins: the marginal social value of an edge's precision and the marginal social value of an edge's existence are the same effective-resistance quantity, differing only in whether $w_e$ is varied at an interior point or pushed from $0$ to $1$. That unification --- a single Laplacian object pricing both which courses students choose and how faithfully instructors grade --- is the organizing observation of the synthesis, and it is visible only because the problem was attacked twice from two directions. Replication, here, was not merely confirmatory; it was generative.

\subsection{Phenomenon 2 (observed): adversarial verification caught real errors}
\label{sec:finding-adversarial}

The second observation is that, in this case, the verification machinery was load-bearing rather than ceremonial. The clearest evidence is P2, the adversarial pair, in which Claude proposes a result and OpenAI Codex (\texttt{gpt-5.5}, high reasoning effort) is tasked with breaking it before the author triages what comes back. Left to its first-pass beliefs, P2 would have shipped three claims that a confident single pass would have presented as theorems. The hostile verifier caught all three. The point here is existence, not rate: adversarial checking demonstrably \emph{caught real errors} on this problem; I do not claim a frequency at which it catches errors in general, which one worked example cannot support. They are worth naming precisely, because they are the point.

\begin{enumerate}
\item \emph{Ex-post incentive neutrality.} P2 initially claimed that the leave-one-student-out (``pivotal'') eigengrade is incentive-neutral ex post, in every realized sample. \methodflaw{This is false.} Codex produced an explicit $K_{2,2}$ counterexample in which a student's realized report shifts the adjustment through the \emph{other} students' shocks; the surviving claim is neutrality \emph{in expectation} only ($\Omega=0$, Prop.~3.1 in the P2 artifact). The error is the subtle kind: true-in-expectation mistaken for true-ex-post.

\item \emph{Budget balance / no-subsidy.} P2 initially asserted a no-subsidy property amounting to budget balance of the instructor transfer. \methodflaw{This is impossible, not merely unproven.} The verifier forced the question to a clean negative: for $m\ge 2$ and $\beta,\eta>0$, exact balance would require the full mixed derivative of $\beta\sum_c \log s_c$ at the target to vanish, but
\[
\partial_{1\cdots m}\Bigl(\beta\textstyle\sum_c \log s_c\Bigr)
\;=\;\beta(-1)^m\, m!\,\eta^m \prod_c s_c
\;=\;\beta(-1)^m\, \frac{m!\,\eta^m}{m^m}\;\neq\;0 ,
\]
the closed form holding when evaluated at the symmetric professional profile $s_c=1/m$, where $\prod_c s_c=1/m^m$.
The corrected artifact (Prop.~4.2) proves balance impossible and offers an AGV expected-externality transfer as the Bayes-Nash fallback, trading dominance for expected balance.

\item \emph{Leave-out necessity for first-moment exactness.} P2 initially claimed the leave-one-out construction was required to make the estimator first-moment incentive-neutral. \methodflaw{The premise was wrong:} the \emph{standard} eigengrade is already first-moment neutral, so the claim was downgraded. What leave-out genuinely buys is \emph{self-influence-freeness} --- $\partial \hat d_c^{(-i)}/\partial \varepsilon_{ic}=0$, a student cannot grade herself up through the adjustment --- a structural Clarke-pivot virtue, not a first-moment one.
\end{enumerate}

None of these errors is sloppy. Each is individually plausible, internally consistent with the surrounding argument, and of precisely the type that survives a careful re-reading by its own author: a cancellation mistaken for balance, an expectation mistaken for a realization, a sufficient construction mistaken for a necessary one. They are the errors a fluent single pass hides best, because fluency is what makes them read as correct.

A scope caveat belongs with this finding. The triage --- deciding which Codex flags were genuine catches and which were false positives (convention mismatches, demands for unnecessary generality) --- was done by one person, the author. I report the three flags that survived triage as real catches because they are demonstrable: each is a named, checkable error left in place in Appendix~\ref{app:methods-output}. I do not report a Codex catch rate or a false-positive rate for this case. The full set of flags Codex raised, and which were accepted or dismissed, is recoverable from the verifier transcripts in the repository, but it was not tallied as a rate, and one worked example would not support such a rate even if it were.

The structured project P3 exhibits the same phenomenon through a different mechanism --- a reviewer gate rather than a counterexample-seeking verifier --- and the example is sharper because the rejected sub-goal had been flagged in advance. P3's literature workstream identified sub-goal SG3, a graph-structured peer-prediction scheme for eliciting honest raw grades, as the component most exposed to prior art and most at risk. When SG3 was developed and submitted to the reviewer gate, it was \emph{rejected}: the leave-one-out residual used to score a grader is independent of that grader's own signal, so the scheme elicits the prior mean and a constant report beats honest reporting. The propriety theorems were false as submitted; the sub-goal is now an explicit honest negative, with its truthfulness claims downgraded to open conjectures and the rejection recorded in the project's decision log. A method whose own gate can reject a sub-goal it spent effort on --- and reject the one it had pre-identified as riskiest --- is doing something a confident single pass structurally cannot.

The numerics carried independent weight in both methods. P2's Monte-Carlo audit confirmed the instructor dominant-strategy claim cleanly (target $\bar G_0+\alpha/\xi=3.40$, best response $3.40$ against every random opponent profile, the Nash alternative matching only at symmetry) while surfacing a precision asymmetry in the leave-out arm that the prose under-discussed. P3's gradient checks (analytic versus numeric agreement to $\sim 10^{-10}$) and its potential-game scan substantiated claims no reader could verify by eye, and its running-example simulation reproduced the Gans--Kominers correlation figures exactly (eigengrade $0.996$ versus letter GPA $0.812$). \methodflaw{P1, by contrast, carried no numerical check of any kind.}

The prescription is correspondingly precise. Separate the prover from the checker, and reward the checker for breaking things. A counterexample-seeking verifier in a read-only sandbox, plus a numerical audit, plus --- in the structured form --- a reviewer gate empowered to reject, caught what proof-writing alone did not. The verification is not a finishing step applied to a finished proof; it is the part of the pipeline that determines which claims are allowed to become proofs at all.

\subsection{Phenomenon 3 (observed): polish is not rigor}
\label{sec:finding-polish}

The third observation is the uncomfortable one, and it is a warning about how AI-assisted theory will be read. Rank the three artifacts by how \emph{finished} they look. P1 is a clean, complete, referee-ready proof document --- the kind of thing one would drop into a paper. P2 is a synthesis document with code, complete on its slice but narrower. P3 is visibly ragged: \unverified{a living \texttt{paper.tex} carrying surviving \texttt{\textbackslash unproven} flags in an appendix}, an explicit open-obligation ledger, one sub-goal marked as an honest negative. By the surface metric of finish, the ordering is P1 $>$ P2 $>$ P3.

Now rank the same three by how \emph{verified} they are --- by counterexample search, numerical cross-check, external review, and honest accounting of gaps. P2 and P3 carry adversarial checks, Monte-Carlo or gradient audits, and explicit flags. P1 carries none: every claim in it is proof-only, with no numerical record whatsoever, and its known limitations are self-flagged by the same pass that produced the proofs. By the metric of verification the ordering is roughly P3 $\approx$ P2 $>$ P1.

\textbf{The most finished-looking output was the least verified.} The most authoritative-looking artifact, P1, is the least independently corroborated, and the most ragged-looking one, P3, is the most disciplined about its own ignorance. With three artifacts I do not read this as a measured correlation between polish and verification; I read it as one clear demonstration that the two come apart, and that the look of finish was, here, the wrong signal to trust. A large language model produces fluent, well-structured LaTeX essentially for free, and that fluency is uniform across claims that are verified and claims that are not. Polish is a property of the \emph{format}; it carries no information about whether the mathematics underneath was checked. A reader who treats a clean proof document as a certificate is reading the one signal the model is best at faking.

The implication for the profession is a shift in how AI-assisted theory must be read and refereed. The relevant question is not whether an artifact looks finished but whether it carries verification infrastructure: counterexample searches, numerical cross-checks, an explicit and current list of open obligations. A proof that \emph{looks} complete is more dangerous than one that advertises its holes, because the former invites the credulity that the latter forecloses. The cultural norm that protects against this is the one P3's strict mode enforces mechanically and at all times: every gap is named. The most trustworthy artifact in the exercise is trustworthy precisely because it tells the reader what it does not know --- and the least trustworthy is so because it does not.

\subsection{A production-function reading: solo, pair, team}
\label{sec:finding-production}

The three methods are recognizably three organizational forms for producing mathematics, and they exhibit the trade-offs an organizational economist would predict. It is worth stating them as such, because the protocol's value is not in any single form but in knowing what each buys.

\begin{description}
\item[Solo expert (P1).] Fastest to a deep, coherent, correct-on-its-slice result, with the cleanest exposition and the sharpest single idea --- here, extensive-margin Pigouvian pricing of enrollments, derived and written referee-ready in one pass. Its weakness is structural: there is no independent check, so blind spots ship as theorems and the absence of numerics is invisible in the output. High output per unit effort; no error correction.

\item[Adversarial pair (P2).] The best reliability \emph{per claim}. The prover/checker split is what caught the three false claims of Section~\ref{sec:finding-adversarial}; in this exercise no other single intervention caught as much. Its weakness is scope: it audited a slice deeply rather than building broadly, and it over-labels its own output (calling a Groves/Pigouvian action-game transfer a ``VCG/Clarke'' mechanism where the literal direct-revelation construction is still aspirational). Maximal trust per claim, narrow coverage.

\item[Structured team (P3).] The best coverage and discipline. It alone produced a literature map (and used it to flag the riskiest sub-goal in advance), a money benchmark, an honest dependency graph across sub-goals, and a reviewer gate that could --- and did --- reject its own work. Its weakness is real but narrower than it first appears: coordination overhead, and a tendency for breadth-of-stubs to \emph{look} like progress before the work is done, so that an interrupted run reads as scaffolding. What completion revealed is that the discipline scales --- multiple sub-goals cleared an adversarial gate with flags, one was honestly killed, and the open-obligation ledger stayed truthful throughout. It is the only one of the three that can reject its own output.
\end{description}

No form dominates. The trade-off is between depth and speed (solo), reliability per claim (pair), and coverage with self-policing (team), and the right choice depends on where in a research project one stands. The natural composition --- and the one the protocol of this paper recommends --- uses each form for what it is best at: a structured team to \emph{map} the problem and maintain the open-obligation ledger, solo provers to \emph{drive} each lemma to a clean statement, and an adversarial verifier gating \emph{every} claim before it is allowed into the paper. The three observations of this section are the motivation for that pipeline. Convergent discovery is why running more than once is worth the cost; the load-bearing role of adversarial verification is why a checker that is rewarded for breaking things is not optional; and the polish--rigor gap is why the ledger of named gaps, not the smoothness of the prose, is the object a reader should trust.

\section{A protocol for practice}
\label{sec:toolbox}

The three runs of Section~\ref{sec:methods} are not competing answers to ``how should one do theory with a language model''; they are three components of one answer, each strong where the others are weak. The multi-agent project maps and bookkeeps but produces breadth that can read as progress before it is verified; the single pass drives a lemma to a clean statement faster than either alternative but ships its blind spots as theorems; the adversarial pair has the highest reliability per claim but, left to itself, audits a slice rather than building a program. The three observations of Section~\ref{sec:findings} turn these complementarities into a workflow: manufacture independence rather than economize on it, treat verification as a gate rather than an afterthought, and track verification status as an explicit object rather than inferring it from prose.

This section states that workflow as a composed pipeline with decision rules for method selection, a fixed set of verification gates, and two artifacts---an open-obligations ledger and a reproducibility trail---that travel with the project from scope to final paper.

\subsection{The composed pipeline}
\label{sec:pipeline}

The default workflow composes the three methods in the order of their comparative strengths: \emph{scope and decompose} with the multi-agent project, \emph{drive each lemma} with the single pass, and \emph{gate every load-bearing claim} with the adversarial verifier and numerics before it enters the paper.

\paragraph{Stage 1: scope and decompose (multi-agent project).} Initialize the problem as a structured project, not a conversation. The scaffolding supplies the four things the other two methods lack at the outset: a living working paper, an explicit research goal, a dependency graph of sub-goals, and---critically---a literature workstream that runs before any proving begins. The map and the prior-art audit are most valuable when they precede the proving, because they determine which lemmas are worth driving and which are likely to collide with known results; in the worked example the literature pass flagged the peer-prediction sub-goal as the riskiest in advance, the same one the reviewer gate later rejected. Output of Stage~1 is a goals file, a sub-goal dependency graph, a literature map, and an initially empty open-obligations ledger with one stub per sub-goal.

\paragraph{Stage 2: drive each lemma to a clean statement (single pass).} For each sub-goal on the critical path, use the single careful pass to convert a target into a precisely stated lemma with a complete candidate proof. Its discipline---state the claim before proving it, sign every term, name rather than bury every gap---produces a statement clean enough to hand to a verifier. The single pass is a \emph{drafting} instrument, not a certifying one: its output is a candidate, and a polished candidate is not a verified one. Output of Stage~2 is, per lemma, a precise statement with explicit hypotheses, a candidate proof, and an honest list of the steps the author is least sure of.

\paragraph{Stage 3: gate the claim (adversarial verifier and numerics).} No claim enters the paper as load-bearing until it has cleared the verification gates of Section~\ref{sec:gates}. The adversarial verifier is a separate model instructed to break the claim---to find a counterexample, a sign error, a missing existence argument, an unjustified step. Its output is treated as a lead, never a verdict: the human triages each finding, distinguishing a real catch (a concrete counterexample, a specific step that does not follow) from a false positive (a convention mismatch, a demand for unnecessary generality, a flag on a stated assumption). Numerical audit runs alongside, because a Monte Carlo or a symbolic check catches what prose hides. A claim that survives the gate is recorded as verified \emph{with its evidence attached}; a claim that fails is either repaired and re-gated or moved to the ledger as open.

The composition is not strictly sequential. A failed gate at Stage~3 sends a lemma back to Stage~2 with a sharper specification, and a counterexample discovered at Stage~3 can force a re-decomposition at Stage~1 (a rejected sub-goal changes the dependency graph). The project scaffolding of Stage~1 is what makes these returns cheap: the ledger and the decisions log record why a claim was downgraded, so the same dead end is not re-explored.

\subsection{Decision rules for method selection}
\label{sec:rules}

The pipeline above is the default for a research program. Not every task warrants the full apparatus, and the comparative strengths in Section~\ref{sec:findings} imply a small set of rules for which method to reach for.

\begin{enumerate}
\item \emph{Use the single pass alone only when the result is small enough to verify by eye.} A short derivation, a sign check, an envelope-theorem step, a one-line monotonicity argument: here the cost of the full gate exceeds its value, because the author can serve as the checker. The moment a result is load-bearing for a downstream claim, it leaves this category and must be gated.
\item \emph{Use the adversarial pair to gate anything load-bearing.} In the worked example the prover/checker split caught three plausible, subtly wrong claims (expectation-truth mistaken for ex-post truth, a cancellation mistaken for budget balance, a construction credited with a property it did not supply) that a confident single pass would have shipped. Any claim a theorem depends on, any claim that will be stated in the paper as established, passes through the adversarial gate.
\item \emph{Use the multi-agent project when the task is a program rather than a lemma.} The project's overhead---coordination, the cost of breadth that can masquerade as progress---is justified when the problem decomposes into many interdependent sub-goals, when a literature map is needed, when the deliverable is a paper with a maintained ledger rather than a single result, and above all when you want a mechanism that can \emph{reject its own work}. The structured team is the only one of the three with a built-in gate that vetoed one of its own sub-goals.
\item \emph{Set the verifier's reasoning effort by the stakes, not the apparent difficulty.} Reserve low effort for routine checks the author is already confident in; use the highest available effort for any unproved load-bearing claim, any counterexample search, and any step the author flagged as uncertain. The cost of high effort is minutes; the cost of a missed counterexample is a false theorem in the paper.
\item \emph{Manufacture independence before trusting a non-obvious derivation.} When a derivation yields a surprising or central object, re-derive it under a different method, a different framing, or a different agent, and check for convergence on the same object. In the worked example two methods independently reached the identical externality kernel from opposite margins; that agreement was the strongest evidence the object was real, and it did mathematical work no single run did, by revealing that one kernel governed both margins. Convergence is cheap corroboration; divergence localizes the bug.
\end{enumerate}

\subsection{Verification gates}
\label{sec:gates}

The protocol's core is a fixed checklist that every load-bearing claim must clear before it is stated as established. The gates are ordered from cheapest to most expensive; a claim that fails an early gate need not consume a later one.

\begin{enumerate}
\item \emph{Self-discipline gate.} The candidate proof states the claim before proving it, signs every term, proves the general case rather than asserting it from examples, cites only standard results, and marks every gap explicitly rather than papering over it with ``clearly'' or ``it is easy to see.'' This gate is enforced by the drafting discipline of Stage~2 and is the minimum bar for a statement to be worth verifying.
\item \emph{Adversarial gate.} A separate model, run at effort matched to the stakes and pointed at the exact statement and proof, attempts to break the claim. The human triages every finding under the lead-not-verdict rule: a real catch is acted on, a false positive is documented and dismissed. A claim passes only after the verifier's confirmed objections are resolved---and ``the verifier found nothing'' is itself recorded as evidence, not assumed.
\item \emph{Numerical gate.} Every claim with computable content is checked numerically against an independent implementation: a Monte Carlo for a distributional claim, a symbolic computation for an algebraic identity, a gradient check for a derivative formula, a direct equilibrium computation for a solution-concept claim. The worked example is emphatic on this point: the most finished-looking output had no numerical check at all and was the least trustworthy, while the numerical audits elsewhere reproduced the source paper's headline figures exactly and caught a precision asymmetry the prose had glossed. A claim with computable content that has no numerical record is not verified, however clean its proof reads.
\item \emph{Reviewer gate.} For results destined for the paper, a final reviewer pass renders an explicit verdict---approve, approve with flags, or reject---and that verdict is written to a durable record rather than left implicit. This gate is what lets a project reject its own work: in the worked example it vetoed a sub-goal whose truthfulness claim did not survive scrutiny, downgrading it to an honest negative. A result that cannot survive a reviewer instructed to reject is not ready for the paper.
\end{enumerate}

A claim is \emph{load-bearing-verified} when it has cleared all four gates and the evidence---the verifier transcript, the numerical script and its output, the reviewer verdict---is attached to it. Anything short of that is open, and open claims live in the ledger, not in the body of the paper as if settled.

\subsection{The open-obligations ledger and the reproducibility trail}
\label{sec:ledger}

Because verification status cannot be inferred from prose, it must be an explicit, maintained object. The protocol carries two such objects through the life of the project.

\paragraph{The open-obligations ledger.} The ledger is a single living list of every claim the project needs but has not yet verified: unproved load-bearing steps, sub-goals not yet attempted, claims downgraded by a failed gate, and modeling simplifications whose relaxation is open. Three rules keep it honest. First, \emph{every gap is named}: an unverified step is marked in place and entered in the ledger when it is identified, never silently carried. Second, \emph{append-and-resolve, not append-and-forget}: when a gate moves a claim from open to verified or from candidate to rejected, the transition and its reason are recorded, so the ledger doubles as a decisions log. Third, \emph{the ledger ships with the paper} as the open-obligations appendix, read as part of the contribution rather than hidden.

\paragraph{The reproducibility trail.} The protocol preserves, for every load-bearing claim, the artifacts behind its verification: the statement and proof, the verifier prompt and its returned transcript (the effort level and which findings were triaged as real versus false positive), the numerical scripts with their seeds and outputs, and the reviewer verdict. Two properties matter. The trail records \emph{negative} results---the counterexamples that killed claims, the sub-goals rejected---because a discarded claim with its refutation is part of the verified frontier. And it records the \emph{human triage}: since the lead-not-verdict rule means the trail is not self-certifying, a reader must be able to audit which verifier findings were accepted and why.

\subsection{The honest deliverable}
\label{sec:deliverable}

It would overstate the case to present this protocol as a recipe for producing finished theorems from current models, and the exercise does not support that claim. None of the three methods proved a hard new theorem from nothing. What the composed pipeline produced---and what a researcher running it today should realistically expect---is a different and still genuinely useful object: a well-mapped problem with a verified core, a clear statement of what is established and by what evidence, and an explicit ledger of what remains open. In the worked example that amounted to a verified externality kernel reached by two routes, a small family of mechanisms with their incentive properties gated and numerically checked, an honest impossibility, an honest negative where a sub-goal failed, and an open-obligations appendix of the steps still to be discharged. The realistic deliverable from current models is not the finished proof but the disciplined frontier, and that is the part of theory-building an unaided blank page makes hardest to do well.

\section{Limitations and threats to validity}
\label{sec:limitations}

The argument of this paper is a methodological one, and it rests on a single experiment. That experiment is informative, but it is narrow in ways I want to state plainly, because the strength of a verification-first protocol is undermined if I am not candid about the conditions under which I observed it work. I group the threats into three: those that limit how far the evidence generalizes, those that concern the status of the three findings as evidence, and those that the protocol itself does not resolve.

\subsection{External validity: one problem, one domain}

The entire comparison runs on a single test problem in a single subfield. I built one VCG-style mechanism for a grading model and ran it through three methods. That problem was chosen deliberately---it is concrete, it has a precisely quantifiable externality handed over by the Gans--Kominers precision bound, and it is hard enough to separate a method that works from one that only looks like it works---but it is still one problem. Mechanism design with a closed-form Laplacian externality is unusually friendly to the convergence I report in Section~\ref{sec:findings}: the externality kernel $\lVert\Pi L^{+}b_{ic}\rVert^{2}$ is a single canonical object, so two independent runs landing on it is less surprising than convergence would be in a domain with many equally natural formulations. Problems whose central object is not so canonical---combinatorics, analysis with delicate constants, areas where the ``right'' definition is itself contested---may show neither the convergent discovery of finding~(1) nor the clean adversarial kills of finding~(2). I make no claim that the three findings transfer beyond the kind of problem demonstrated here, and a reader should treat this as a worked existence proof for the protocol, not as a characterization of its reach.

\subsection{Model-version dependence}

Every result reported here was produced by Claude Opus 4.8, with the adversarial method pairing it against OpenAI Codex (gpt-5.5) at high reasoning effort. The protocol is designed to be model-agnostic---it is a discipline for verification, not a property of any particular model---but the \emph{outputs} I observed, and therefore the three findings, are tied to these specific versions. Which false claims a single pass ships, whether the adversary finds the $K_{2,2}$ counterexample to ex-post neutrality, whether the reviewer gate rejects the peer-prediction sub-goal rather than waving it through: all of these are behaviors of particular model weights at a particular moment. A stronger model might make fewer of the errors that make the adversarial step look load-bearing; a weaker one might produce kills that are themselves wrong. I cannot promise that re-running this protocol on later models, or on a different proposer/verifier pairing, reproduces the same pattern, and I would expect the specific error inventory in Appendix~\ref{app:methods-output} to be the least stable part of the paper across versions.

\subsection{A structured case study, not a randomized comparison}

The three methods were not run as a controlled experiment. I did not randomize problems across methods, fix a common compute or wall-clock budget, hold the prompt sequence constant, or pre-register the outcomes I would count as success. The three runs differ on the margin they attacked (student versus instructor), the instrument they used (money versus report space), and the solution concept they reached, in addition to differing on how the work was verified---so the comparison confounds the verification axis I care about with several others. This means the three findings are \emph{demonstrations}, not estimates. ``Adversarial verification is load-bearing'' is established by exhibiting three real false claims that the adversary killed (Section~\ref{subsec:adversarial}); it is not an estimated rate at which adversarial verification catches errors, and nothing here licenses a quantitative claim of that form. ``Polish is not rigor'' is an observed inversion on three data points, not a measured correlation. The honest reading is that this paper shows these phenomena \emph{can} occur and gives mechanism for \emph{why} they occur, while leaving their frequency unmeasured.

\subsection{The theory is partly open}

The worked example is a research program with a verified core, not a finished theory, and several of its steps remain unproven. The multi-agent project's peer-prediction sub-goal (SG3) was rejected at the reviewer gate---the leave-one-out residual is independent of the grader's own signal, so it elicits the prior mean and a constant report beats honest reporting---and stands as an explicit negative rather than a resolved component. The student-side mechanism (SG2), the no-money impossibility wall (SG4), and the composite two-sided existence theorem (SG5) carry conditions and gaps that the strict-mode flags record but do not discharge; the SG1-a potential-game existence claim survives only after the reviewer forced it back from a global statement to a coercivity/compact-box condition. The single-pass mechanism's clean individual-rationality proof holds for the course-contrast target and breaks for the student-ability target, and it received no numerical verification of any kind. I report these not as failures of the protocol but as its intended output: the surviving \unverified{} flags in Appendix~\ref{app:methods-output} are the ledger of what the methods did \emph{not} establish, and the reader should weigh the worked example accordingly.

\subsection{Verifier false positives and false negatives}

The protocol leans on a verifier, and a verifier is fallible in both directions. The adversarial method treats every Codex finding as a lead and not a verdict precisely because the verifier is a strong but unreliable mathematician: it raises false positives---convention mismatches, demands for unnecessary generality---that a human must triage, and triage is itself a point of failure I did not measure. The more dangerous direction is the false negative. A verifier that fails to break a claim does not certify it; it only fails to refute it on the attempts made. The reassurance I draw from convergent discovery and from surviving adversarial attack is evidence, not proof, and a coordinated error---a flawed lemma that both the proposer and the verifier share, or a numerical check that confirms a claim only on the single seed it was run with---would pass through this protocol undetected. The Monte Carlo audits in the adversarial run were single-seed, and at least one of them showed a second-moment precision asymmetry that the writeup under-discussed; that is exactly the kind of artifact a thinly-sampled check can hide.

\subsection{Selection in what I chose to report}

Finally, I am the author, the operator, and the triager, and the paper reflects my choices about what to surface. I decided which claims to escalate to the adversary, which of its findings were real catches, which runs to present, and which details from each writeup to foreground. The error inventory in Appendix~\ref{app:methods-output} is the set of mistakes that were \emph{caught and recorded}; there is no way for me to report the errors that survived all three methods and my own reading, and a clean-looking ledger may reflect the limits of the checking as much as the quality of the work. The protocol is designed to reduce this selection pressure---the strict-mode gate records rejections I would otherwise be tempted to bury, and the adversarial split moves the catch off my own judgment---but it does not eliminate it.

\subsection{Authorship and reproducibility}

When the substantive results are machine-generated, two further questions arise that the protocol does not settle. The first is authorship: the derivations, counterexamples, and much of the prose in the three writeups were produced by the models, with my role being to scope the problem, run the methods, triage the verifier, and judge the output. I have tried to be exact about provenance throughout---each result in the worked example is attributable to a specific method---but the line between authoring a result and operating a system that produces it is one this paper does not pretend to draw cleanly, and norms for crediting and refereeing such work are not yet settled. The second is reproducibility. The skills, the example outputs, and the artifacts are released openly so that the \emph{protocol} is reusable and inspectable.\footnote{Repository: \url{https://github.com/morankor/theorist-toolbox} (to be made public).} But the underlying models are accessed through commercial interfaces, are not version-pinned in a way I control, and are stochastic; another operator running the same skills on the same problem should expect outputs that rhyme with mine rather than reproduce them exactly. What is meant to be reproducible here is the discipline and the released artifacts, not the token-level trajectory that produced them.

\section{Conclusion}
\label{sec:conclusion}

Empirical economists inherit craft; theorists inherit a blank page. The asymmetry I set out to address is not that theorists lack a capable model---by 2026 the models prove---but that they lack a discipline for trusting what the model proves. The contribution of this paper is that discipline: a verification-first protocol for doing economic theory with a language model, instantiated as three reusable methods that differ on a single axis, how the work is checked. The single pass commits to a fixed proof discipline and nothing more. The adversarial pair sets one model to propose and a second to break, treating each adversarial finding as a lead rather than a verdict, with Monte Carlo checks alongside. The multi-agent project scaffolds the proof the way one runs a research team, with separate workstreams, a strict-mode obligations ledger, and a reviewer gate empowered to reject. The grading mechanism was the worked example, chosen because it is hard enough to separate a method that works from one that only looks like it works; the comparison across the three methods is the result.

I want to be plain about where the credibility of this protocol lives. It is not in the polish of the output. The central lesson of the three-method comparison is that polish and rigor come apart, and that they invert: the most finished-looking artifact, the single pass, carried no numerical check and was the least verified, while the raggedest, the multi-agent suite covered in its own open-obligation flags, was the most trustworthy precisely because it stated what it did not know. Credibility lives instead in the scaffolding that the protocol makes load-bearing---the adversarial gate that killed three of my own false claims, the numerics that reproduced the source paper's headline figures and matched closed forms to ten digits, and the open-obligations ledger that left a rejected sub-goal marked unproven rather than quietly dropped. When two of the methods, run separately, derived the same externality kernel from opposite margins, that agreement was cheap replication and I believed the object. None of these are properties of the prose. They are properties of the process, and the right thing to read for in machine-assisted theory is the verification record, not the writing.

I am equally plain about the ceiling. None of this proved a hard new theorem from nothing. What the three methods produced together is a well-mapped problem with a verified core, a genuine impossibility result, a working money-free mechanism, and an explicit ledger of what remains open---the realistic deliverable from a 2026 model, and a useful one. As the models improve, I expect the binding constraint to remain where it is. Better models will close more of the open obligations and start more proofs from a fuller page, but a stronger prover is also a more persuasive source of subtle, plausible error, and the value of an adversarial gate, a numerical check, and a ledger that refuses to mark a thing done rises rather than falls with the fluency of the thing being checked. The protocol is built to outlast the model it runs on: its commitments are to verification, not to any particular system's competence. That is why I wrote a paper and not only the code. A theorist should not have to start every proof from a blank page, and should not have to take a machine's proof on faith either; the toolbox is an attempt to remove the first burden without incurring the second.

\appendix

\section{The methods' output, reproduced as is}
\label{app:methods-output}

This appendix reproduces the mathematical output of the three methods of
Section~\ref{sec:methods} essentially as each produced it, including its
mistakes. The purpose is to let the reader judge the methods directly rather
than through my paraphrase: a method's value is partly in what it gets right and
partly in what it gets wrong, and a corrected reproduction would erase exactly
the evidence I want to display. I have therefore condensed definitions, theorem
statements, and proofs for length, but I have \emph{not} repaired any
substantive error, overclaim, or gap. Where a claim is known to be false,
incomplete, or overstated --- as established either by the method's own internal
checking (Method~2's adversary, Method~3's reviewer gate) or by my triage --- I
mark the spot with a \methodflaw{} note that states what is wrong and how it was
caught, and I leave the flawed claim itself in place. Every \methodflaw{}
error-flag in this appendix was re-verified against the source artifacts before
publication: each flag was checked back against the run that produced it and
against the verifier transcript or reviewer record that caught it, so the flags
report what the methods actually established. Read every
\methodflaw{} as an editorial intrusion that was \emph{not} present in the
source artifact; the corresponding example outputs in the repository
(the single-pass enrollment writeup, the adversarial-pair grading writeup, and the
multi-agent \texttt{paper.tex}) contain the unmarked claims.

Notation across the three subsections is each method's own and is not
reconciled; in particular ``$L$'' is the bipartite (student--course) signed
Laplacian throughout, $b_{ic}$ is the signed incidence vector of edge $(i,c)$,
$\Pi$ (or the centered projector $M_n$) projects onto the student-contrast
space, and $\sigma^2$ is the grading-noise variance, all inherited from
\citet{ganskominers2026}.

\subsection{Method 1 (single pass): a student-side enrollment Groves mechanism}
\label{app:m1}

The single disciplined Claude pass produced a Groves/Pigouvian mechanism on the
\emph{extensive} margin: it prices which courses a student takes, taking the
neutrality of eigengrades (the strategic susceptibility $\Omega(\mathcal R)=0$ of
\citealp{ganskominers2026}) as the starting point and adding the one missing
incentive --- a Pigouvian subsidy equal to the marginal social informativeness of
each enrollment.

\paragraph{Setup.} Students $S=\{1,\dots,n\}$, courses $C=\{1,\dots,m\}$,
$N=n+m$. Latent performance $P_{ic}=a_i-d_c+\varepsilon_{ic}$ with
$\operatorname{Var}(\varepsilon_{ic})=\sigma^2$, stacked as $x^\star=(a,d)$. An
enrollment graph $\mathcal G=(S\cup C,E)$ yields signed incidence vectors $b_e$
($b_e^\top x^\star=a_i-d_c$ for $e=(i,c)$), incidence matrix $B$, and Laplacian
$L=B^\top B=\sum_e b_e b_e^\top$; the observation model is $p=Bx^\star+\varepsilon$.
For a target contrast matrix $K$ whose rows lie in $\mathbf k^\perp$ (with
$\mathbf k=\tfrac1{\sqrt N}(\mathbf 1_n,\mathbf 1_m)\in\ker L$), the
\emph{informational cost} of a connected graph is
\[
  \Psi(E)\;=\;\operatorname{tr}\!\big(K\,L(E)^{+}K^\top\big),
\]
and $\Psi(E)=+\infty$ when $E$ is disconnected. Two targets are distinguished:
\textbf{(T-course)}, the centered course effects $d-\bar d\,\mathbf 1_m$ (fixed
dimension $m$), taken as the leading specification; and \textbf{(T-stud)}, the
centered student effects $a-\bar a\,\mathbf 1_n$, for which (with $K$ the centered
projector) $\Psi(E)=\operatorname{tr}(L^{+})$ reproduces the precision bound
$\sigma^2\operatorname{tr}(L^+)\le\sigma^2(n+m-1)/\lambda_2(L)$ of
\citet[Prop.~6.15]{ganskominers2026}.

Each student $i$ has private pedagogical values $v_i=(v_{ic})_{c}\in\mathbb
R_{\ge0}^m$ and quasilinear utility $u_i=v_i(E)+t_i$. The institution (``agent
$0$'') has public valuation $V_0(E)=-\lambda\sigma^2\Psi(E)$, $\lambda>0$, so
social surplus is $W(E)=\sum_{j\in S}v_j(E)+V_0(E)$. The direct mechanism
collects reports $\hat v$, chooses $E^\star(\hat v)\in\arg\max_{E\in\mathcal
F}\widehat W(E)$, and pays the Groves transfer
\[
  t_i(\hat v)=\sum_{j\ne i}\hat v_j\!\big(E^\star(\hat v)\big)+V_0\!\big(E^\star(\hat v)\big)-h_i(\hat v_{-i}),
\]
with the Clarke pivot $h_i(\hat v_{-i})=\mathcal W(S\setminus i)$, the optimal
welfare of the economy without $i$.

\paragraph{The informational-cost lemma.} The mechanism rests on a least-norm
characterization of BLUE variance: for any estimable contrast $k^\top x^\star$,
$\operatorname{Var}(\widehat{k^\top x^\star})=\sigma^2\,\nu(k;B)$ where
$\nu(k;B)=\min\{\|\ell\|^2:B^\top\ell=k\}$.

\begin{quote}
\textbf{Lemma~1 (properties of $\Psi$).}
(a) BLUE variance equals $\sigma^2\nu(k;B)$.
(b) \emph{Edge-monotonicity:} adding one observation weakly lowers $\Psi$.
(c) \emph{Participation monotonicity (target T-course):} adding a student's edges
weakly lowers $\Psi$.
(d) \emph{Convexity:} on intensities $z\ge0$ with $L(z)=\sum_e z_e b_e b_e^\top$,
the map $z\mapsto\operatorname{tr}(KL(z)^+K^\top)$ is convex.
\end{quote}

The proof of (b) splits any $\ell'=(\ell,\ell_{e'})$ feasible for the augmented
problem $B'^\top\ell'=k$; since $(\ell_0,0)$ is feasible whenever $B^\top\ell_0=k$
and preserves the norm, $\nu(k;B')\le\nu(k;B)$, and summing over the rows of $K$
gives $\Psi(E\cup\{e'\})\le\Psi(E)$. Part (c) applies (b) repeatedly to add a
student's edges, using that a (T-course) contrast $k$ has $k_i=0$ so it is
unchanged across the $S$- and $(S\setminus i)$-economies. Part (d) writes
$\Psi(z)=\operatorname{tr}(KA(z)^{-1}K^\top)$ with $A(z)=\Pi L(z)\Pi$ affine and
shows the second derivative along $A(s)=A+sH$ equals
$2\big\|SA^{-1/2}K^\top\big\|_F^2\ge0$ with $S=A^{-1/2}HA^{-1/2}$. Differentiating
a single intensity yields the marginal informativeness identity
\begin{equation}\label{eq:m1-marginal}
  \frac{\partial\Psi}{\partial z_e}=-\big\|K L^{+}b_e\big\|^2\;\le0,
\end{equation}
which for (T-stud) becomes $-\|L^+b_e\|^2$, a squared effective resistance.

\paragraph{Main results.} The mechanism's theorems are the standard Groves
conclusions, instantiated for this objective.

\begin{quote}
\textbf{Theorem~2 (dominant-strategy truthfulness).} With any Groves pivot,
truthful reporting $\hat v_i=v_i$ is weakly dominant for every student.
\end{quote}

\noindent The proof defines $\Phi(E)=v_i(E)+\sum_{j\ne i}\hat v_j(E)+V_0(E)$;
since $h_i$ is independent of $\hat v_i$, student $i$'s utility is
$\Phi(E^\star(\hat v_i,\hat v_{-i}))-h_i$, and truthful reporting makes
$\widehat W\equiv\Phi$, so the induced graph maximizes $\Phi$ over $\mathcal F$.

\begin{quote}
\textbf{Theorem~3 (first-best curriculum).} In the truthful equilibrium,
$E^\star(v)\in\arg\max_{E\in\mathcal F}W(E)$; edge $(i,c)$ is included iff
$v_{ic}\ge\lambda\sigma^2[\Psi(E^\star\setminus\{(i,c)\})-\Psi(E^\star)]$.
\end{quote}

\begin{quote}
\textbf{Theorem~4 (individual rationality, target T-course).} Under the Clarke
pivot and the free-entry condition that the $(S\setminus i)$-economy's optimal
graph remains feasible after adding $i$'s edges, $u_i=\mathcal W(S)-\mathcal
W(S\setminus i)\ge0$ for every student.
\end{quote}

\noindent The IR proof shows $\mathcal W(S)\ge\mathcal W(S\setminus i)$ by
exhibiting a feasible $S$-economy graph $E'$ extending the
$(S\setminus i)$-optimum: the other students' pedagogical sum is unchanged,
$v_i(E')\ge0$, and by Lemma~1(c) the course-contrast cost does not rise,
$\Psi_S(E')\le\Psi_{S\setminus i}(E^\star_{-i})$.
\methodflaw{The IR argument is sound \emph{only} for the (T-course) target,
because it invokes Lemma~1(c), participation monotonicity, which the source
itself proves only for the fixed-dimension course contrast. For the
(T-stud) student-ability target the source's own Remark after Lemma~1(c) concedes
that adding a student introduces a \emph{new} contrast $a_i-\bar a$ whose variance
enters $\Psi$, so $\Psi_S(E')\le\Psi_{S\setminus i}(\cdot)$ ``need not hold''
and the clean IR conclusion $u_i\ge0$ breaks; IR for (T-stud) is left as an
unverified explicit condition (``$i$'s marginal contribution \dots\ minus her own
estimation variance is nonnegative''). Thus the headline target a reader most
naturally wants --- recovering student ability --- is exactly the one for which
the IR theorem is not established.}

\begin{quote}
\textbf{Proposition~6 (Pigouvian characterization).} Holding others fixed,
$\partial u_i/\partial(\text{add }(i,c))=v_{ic}-\lambda\sigma^2[\Psi(E)-\Psi(E\setminus\{(i,c)\})]$,
a subsidy equal to the true marginal social informativeness; in the intensity
relaxation it is $\lambda\sigma^2\|KL^+b_{ic}\|^2\ge0$ by~\eqref{eq:m1-marginal}.
The transfer's course-dependence carries \emph{no} $-d_c$ arbitrage term, because
$\Psi$ depends on $E$ only through $L=\sum_e b_eb_e^\top$, built from incidence
vectors and not from realized course effects.
\end{quote}

\begin{quote}
\textbf{Proposition~7 (welfare dominance).}
$W(E^\star)\ge W(E^{\mathrm{ped}})$ over the pedagogically-optimal (neutral)
curriculum, strictly whenever pedagogy and informativeness disagree.
\end{quote}

The pass closes with remarks: a convex (A-optimal experimental-design)
relaxation $\max_z\sum_e z_e v_e-\lambda\sigma^2\operatorname{tr}(KL(z)^+K^\top)$
solvable to global optimality; the observation that the Clarke pivot is
\emph{not} budget-balanced, with the deficit attributed to the
Green--Laffont/Hurwicz impossibility for a public-good objective on a connected
type domain; and a scope note (it does not address instructor honesty, collusion,
or false-name enrollment).

\methodflaw{The entire Method~1 artifact contains \emph{no numerical
verification whatsoever} --- no Monte Carlo, no worked instance, not even the
paper's own 5-student/8-course running example. Every claim above is
analytical only. This is the central limitation of the single-pass method: the
output is the most polished and internally consistent of the three (clean Groves
structure, correct least-norm proofs), yet it is the least \emph{checked}. The
unverified-IR gap for (T-stud) flagged above is precisely the kind of
target-dependent failure a numerical probe on a concrete graph would have surfaced
immediately, and did not, because no such probe was run.}

\subsection{Method 2 (adversarial pair): an instructor-side Pigouvian transfer}
\label{app:m2}

The adversarial method (Claude proposing, OpenAI Codex/gpt-5.5 at high effort
attempting to break each claim, author triage, plus NumPy Monte Carlo) produced
an \emph{instructor}-side mechanism that prices grade inflation while keeping raw
grades, together with a sharper reading of eigengrades as a Groves-form
correction. I reproduce the four substantive components and flag the three claims
that the adversary killed --- which are the point of displaying this method.

\paragraph{Eigengrades as a Groves-form correction (\S2--\S3 of the source).}
Writing a linear full-matrix score $T_i(P)=\sum_{j,c}L_{i,jc}P_{jc}$, the source
restates \citet[Prop.~6.3]{ganskominers2026} as an annihilator characterization
(Prop.~2.1): the scores invariant to every additive course shift
$P\mapsto P-\mathbf 1_n h^\top$ are exactly the arrays with $\sum_j L_{i,jc}=0$
for each $c$, and among these the ones returning $a_i-\bar a$ are those that also
satisfy $\sum_c L_{i,jc}=\mathbf 1\{j=i\}-1/n$. The ``Groves reading'' is that
the own-grade part of $T_i$ is corrected by a reference term $H_i(P_{-i})$ built
from \emph{others'} data --- structurally the Clarke-pivot idea of cancelling a
contaminating common component using information the agent does not control. The
source is careful that this is an analogy, not an equivalence (no money, no
allocation, no quasilinear payoff identity).

The \emph{pivotal (leave-one-out) eigengrade} makes the pivot literal: deleting
student $i$ and her edges, compute on $\mathcal G_{-i}$ the course-block estimate
$\hat d^{(-i)}$ (a function of $\{P_{jc}:j\ne i\}$ only) and report
$Y_i(c)=P_{ic}+\hat d_c^{(-i)}$.

\begin{quote}
\textbf{Proposition~3.1 (first-moment incentive-neutrality, exact).} If
$\mathcal G_{-i}$ is connected, then
$m_i(c;\mathcal R)=\mathbb E[Y_i(c)\mid a,d,\,i\text{ chooses }c]=a_i-\kappa_{-i}$
for every feasible $c$, with $\kappa_{-i}$ independent of $i$'s choice; hence
$\Omega(\mathcal R)=0$ exactly, and under an external anchor $\kappa_{-i}=0$.
\end{quote}

Two clarifications were attached to Proposition~3.1 in the source.

\methodflaw{Claim (a), \emph{ex-post} incentive neutrality. The source's first
draft asserted that first-moment neutrality holds ``ex-post in every realized
sample,'' not merely in expectation. The adversary (Codex) \emph{disproved} this
with a $K_{2,2}$ / $n=3$ counterexample: the realized \emph{other-student} shocks
make any single transcript sample-dependent, so the leave-out adjustment is
choice-invariant only in expectation. The claim was corrected in place to ``in
expectation,'' and the residual ex-post channel is acknowledged to be a
second-moment (precision) effect under a Bayesian market. The verification table
records this as ``My `ex-post in every sample' claim was \textbf{false} (Codex
$n=3$ counterexample) --- corrected to `in expectation.'\,''}

\methodflaw{Claim (c), the \emph{necessity} of leave-out for first-moment
exactness. The source originally premised that leave-one-out was needed to obtain
first-moment neutrality. Monte Carlo plus Codex showed the \emph{standard}
(non-leave-out) eigengrade is \emph{already} first-moment neutral in the
unconditional mean (a null effect, $z=0.27$); the original premise was therefore
\emph{wrong} and was \emph{downgraded}. What the leave-out estimator genuinely
buys is not first-moment exactness but \emph{self-influence-freeness}: in the
standard estimator student $i$'s own shock enters $\hat d_c$ with weight $-1/N_c$
($\partial\hat d_c/\partial\varepsilon_{ic}=-1/N_c\ne0$), and leave-out sets this
to exactly zero --- a structural Clarke-pivot virtue, not a first-moment one.}

Proposition~3.2 prices the pivot: under zero-sum leave-out normalization
$\kappa_{-i}=\bar a_{-i}=(A-a_i)/(n-1)$, giving an $O(1/n)$ centering bias
$(m_i-m_j)-(a_i-a_j)=(a_i-a_j)/(n-1)$, removed by an external anchor or the
rescaling $\tilde Y_i=\tfrac{n-1}{n}Y_i$.

\paragraph{The Groves/Pigouvian transfer (\S4 of the source).} This is the method's
main constructive deliverable, a corrective transfer over instructors' grade
actions (not a direct-revelation VCG/Clarke mechanism), targeting the instructor game of
\citet[\S8]{ganskominers2026}. Each instructor $c$ chooses mean grade $\bar G_c$;
logit enrollment shares $s_c(\bar G)=e^{\eta\bar G_c}/\sum_k e^{\eta\bar G_k}$
($\eta=\gamma\rho_{\mathcal R}\ge0$, with $\eta=0$ under exact eigengrades);
intrinsic payoff
\[
  U_c(\bar G)=\alpha\bar G_c+\beta\log s_c(\bar G)-\tfrac{\xi}{2}(\bar G_c-\bar G_0)^2 .
\]
The unique symmetric equilibrium is
$\bar G^{\mathrm{NE}}=\bar G_0+(\alpha+\beta\eta(1-1/m))/\xi$, whose term
$\beta\eta(1-1/m)/\xi$ is the competitive-enrollment (business-stealing) component
of inflation; the planner values only $W(\bar G)=\sum_k[\alpha\bar
G_k-\tfrac\xi2(\bar G_k-\bar G_0)^2]$, maximized at the professional profile
$\bar G_c=\bar G_0+\alpha/\xi$. The mechanism keeps raw grades and pays
\[
  t_c(\bar G)=-\beta\log s_c(\bar G)+h_c(\bar G_{-c}),
  \qquad h_c\ \text{a function of others' grades only}.
\]

\begin{quote}
\textbf{Theorem~4.1 (Groves-form, dominant-strategy efficiency).}
(i) The $\beta\log s_c$ terms cancel: $U_c+t_c=[\alpha\bar G_c-\tfrac\xi2(\bar
G_c-\bar G_0)^2]+h_c(\bar G_{-c})$. (ii) The residual objective is strictly
concave with maximizer $\bar G_c^\star=\bar G_0+\alpha/\xi$ independent of $\bar
G_{-c}$, $\eta$, and $\beta$; the professional profile is the unique strictly
dominant-strategy equilibrium \emph{even when $\eta>0$}, with raw grades retained.
(iii) The implemented outcome equals the $\eta=0$ eigengrade equilibrium of their
Cor.~8.4.
\end{quote}

\noindent Dominance follows because after cancellation the objective is separable,
hence opponent-independent; the source contrasts the naive ``internalize-others''
transfer $\tilde t_c=\beta\sum_{k\ne c}\log s_k$, whose FOC
$\alpha-\xi(\bar G_c-\bar G_0)+\beta\eta(1-ms_c)=0$ hits the target only at the
symmetric profile $s_c=1/m$ (Nash, not dominant). Theorem~4.1(i)--(iii) was
verified line-by-line by Codex and by Monte Carlo, and is among the method's
\emph{surviving} claims.

\methodflaw{Claim (b), budget balance / no-subsidy. The source's original
Theorem~4.1(iv) claimed the scheme could be made no-subsidy / budget-balanced.
Adversarial checking proved this \emph{false} and the claim was \emph{removed},
replaced by the honest Proposition~4.2 below. Two facts kill it: (1) since
$\log s_c\to-\infty$ as $\bar G_c\to-\infty$, the transfer becomes a subsidy on
the unbounded grade domain, so no Clarke pivot $h_c$ makes $t_c\le0$ everywhere;
and (2) exact budget balance $\sum_c t_c=0$ is \emph{impossible} under smooth
dominant-strategy implementation whenever $m\ge2,\beta>0,\eta>0$, because balance
would force the full mixed derivative $\partial_{1\cdots m}\sum_c U_c$ to vanish
at the target, yet
\[
  \partial_{1\cdots m}\Big(\beta\sum_c\log s_c\Big)
   =\beta(-1)^m m!\,\eta^m\prod_c s_c
   =\beta(-1)^m m!\,\eta^m/m^m\ \ne\ 0 .
\]
The fallback is an AGV / expected-externality transfer, which restores budget
balance only \emph{in expectation} and only under \emph{Bayes--Nash} (not
dominant-strategy) implementation. This is the textbook VCG budget tension, and
the method's value here is that the adversary forced the honest version
(Prop.~4.2) in place of the overclaim.}

\begin{quote}
\textbf{Proposition~4.2 (budget --- the VCG obstruction, honest).} On the
unbounded domain $t_c\le0$ cannot be guaranteed; exact balance is impossible for
$m\ge2,\beta,\eta>0$ by the mixed-derivative identity above; AGV restores balance
in expectation at the cost of moving to Bayes--Nash; on a bounded grade interval
the target is the projection of $\bar G_0+\alpha/\xi$ and a finite entry fee
makes the scheme ex-post individually rational.
\end{quote}

The artifact closes with an explicit verification table (Codex high-effort plus
Monte Carlo, all findings triaged) listing each claim with verdict ---
Prop.~2.1 confirmed, ``eigengrades \emph{are} Groves'' downgraded to analogy,
Prop.~3.1 confirmed in expectation with the ex-post claim corrected, the
standard-estimator first-moment exactness confirmed (premise corrected),
Prop.~3.2 confirmed, Thm.~4.1(i)--(iii) confirmed, and (iv) ``\textbf{FALSE ---
removed},'' replaced by Prop.~4.2. The Monte Carlo scripts and Codex transcripts
are retained as artifacts (\texttt{output/leaveout\_montecarlo.py},
\texttt{output/vcg\_instructor\_check.py}, \texttt{output/codex\_explorations/}).

\subsection{Method 3 (multi-agent project): a money-free report-space suite}
\label{app:m3}

The multi-agent method (architecture after \citet{zheng2026}; strict mode,
workstreams, a reviewer gate) produced a suite of report-space mechanisms that
use \emph{no} monetary transfers, internalizing the eigengrade-precision
externality with fixed-supply numéraires (enrollment shares, priority points)
instead of money. Its central object is the externality kernel, derived
independently of Method~1 on the \emph{intensive} margin.

\paragraph{Environment and the externality kernel.} Instructor $c$ injects
within-course reporting noise of variance $\tau_c^2\ge0$, reporting
$g_{ic}=P_{ic}+h_c+\zeta_{ic}$, $\zeta_{ic}\sim(0,\tau_c^2)$; the institution runs
weighted GLS with edge precisions $w_{ic}=(\sigma^2+\tau_c^2)^{-1}$, $L_W=B^\top
WB$, $\hat x=L_W^+B^\top Wp$, so $\operatorname{Cov}(\hat x)=L_W^+$
(Lemma~\ref{app:m1}-style GLS identity). With $\Pi=\operatorname{diag}(M_n,0_m)$,
the informativeness welfare is
$W(\tau)=-\operatorname{tr}(\Pi L_W^+\Pi)$.

\begin{quote}
\textbf{Proposition (edge-precision gradient).} For every edge $e$,
$\partial W/\partial w_e=\|\Pi L_W^+ b_e\|^2\ge0$, and hence
\[
  \frac{\partial W}{\partial\tau_c}
   =-\frac{2\tau_c}{(\sigma^2+\tau_c^2)^2}\sum_{i:(i,c)\in E}\|\Pi L_W^+ b_{ic}\|^2\;\le0 .
\]
\end{quote}

\noindent The proof uses $b_e^\top\mathbf 1=0$ to kill the
pseudoinverse-correction terms in the Golub--Pereyra differential, leaving
$\partial L_W^+=-L_W^+ b_eb_e^\top L_W^+$. The quantity $\|\Pi L_W^+ b_{ic}\|^2$
is the effective-resistance-type \emph{measurement-precision externality} of
course $c$'s compression --- the \emph{same} kernel Method~1 derived on the
extensive margin (cf.~\eqref{eq:m1-marginal} with $K=\Pi$).

A money-VCG benchmark (SG0) establishes that, with Groves transfers $t_c$ paid in
a transferable numéraire, truthful reporting of the instructor's compression
taste $\theta_c$ is dominant and the efficient rule sets $\tau_c=0$
(full fidelity). The well-posedness analysis notes that $F(w)=\operatorname{tr}(\Pi
L_W^+\Pi)$ is convex in the edge-precision vector $w$ (operator convexity of the
inverse) but that, written in compression, $W_\tau(\tau)$ is concave only on the
low-compression box $\{0\le\tau_c\le\sigma/\sqrt3\}$ and fails to be concave on
$[0,\infty)^m$ (an explicit $K_{2,2}$ instance is given).

\paragraph{The instructor mechanism (SG1) and the simplex tax.} The numéraire is
enrollment share, fixed-supply ($\sum_c s_c=1$); the institution sets
$s_c(\tau)=e^{-\eta\psi_c(\tau_c)}/\sum_k e^{-\eta\psi_k(\tau_k)}$ and instructor
$c$ has payoff $V_c(\tau)=u_c(\tau_c)+\beta\log s_c(\tau)$.

\begin{quote}
\textbf{Proposition (share-pivot reproduces the money-pivot, taxed by $1-1/m$).}
The FOC is $u_c'(\tau_c)=\beta\eta\psi_c'(\tau_c)(1-s_c)$; in a course-symmetric
environment the efficient symmetric profile $\tau^\star\mathbf 1$ is a Nash
equilibrium at intensity $\eta^\star=u'(\tau^\star)/[\beta(1-1/m)\psi'(\tau^\star)]$,
matching the money-pivot marginal $-\partial_{\tau_c}W(\tau^\star\mathbf 1)$.
\end{quote}

\noindent The factor $1-1/m$ --- the ``simplex tax'' --- is the zero-sum cost of a
fixed-supply numéraire, and is identified with the same $1-1/m$ in
\citet{ganskominers2026}'s competitive-inflation term $\beta\eta_{\mathcal
R}(1-1/m)/\xi$ (their eq.~19). The global structure is then analyzed (SG1-a): the
share game has an \emph{exact potential}
$\Phi(\tau)=\sum_c[u_c(\tau_c)-\beta\eta\psi_c(\tau_c)]-\beta\Lambda(\tau)$ with
$\Lambda=\log\sum_k e^{-\eta\psi_k(\tau_k)}$, verified by
$\partial_{\tau_c}\Phi=\partial_{\tau_c}V_c=u_c'-\beta\eta\psi_c'(1-s_c)$; the
Hessian decomposes as $\operatorname{diag}(u_c'')-\beta\eta\operatorname{diag}(\psi_c''(1-s_c))
-\beta D(\operatorname{diag}(s)-ss^\top)D$, and under ``$u_c$ concave, $\psi_c$
convex'' each block is signed so $\Phi$ is concave.

\begin{quote}
\textbf{Theorem (concavity, global best response, uniqueness).} Under those
hypotheses $\Phi$ is concave (strictly if $u_c''<0$); each $V_c$ is own-concave so
the FOC characterizes the global best response; and the equilibria are exactly the
maximizers of $\Phi$, unique under strict concavity \emph{when one exists}.
\end{quote}

\methodflaw{SG1-a \emph{existence} overclaim. The status line and the first
iteration of SG1-a presented the share game as delivering a \emph{unique
equilibrium} for arbitrary heterogeneous instructors --- i.e.\ implicitly claimed
existence. The reviewer gate produced a counterexample forcing existence back to a
\emph{coercivity / compact-box} hypothesis: with $m=2$, $\beta=\eta=1$,
$u_c(t)=2t$ (concave) and $\psi_c(t)=0.1\,t$ (convex), the potential
$\Phi(t,t)=1.9t-\beta\Lambda$ \emph{diverges} to $+\infty$ along the diagonal,
each $V_c$ is strictly increasing in its own argument, and the share game has
\emph{no} Nash equilibrium on $[0,\infty)^m$. The theorem as reproduced therefore
gives concavity, global best response, and \emph{at-most-one} equilibrium
unconditionally, but existence \emph{only} under either a compact action box
$\tau_c\in[0,\bar\tau]$ (Assumption~A-coerc) or a coercivity condition
$u_c(t)-\beta\eta\psi_c(t)\to-\infty$ (H-coerc). The qualifier ``when it exists''
in the theorem statement is the residue of this correction; the unconditional
existence claim was withdrawn.}

The student mechanism (SG2) prices the \emph{extensive} margin with a fixed-budget
Hylland--Zeckhauser/Budish pseudo-market: equal priority-point budgets, seat
prices $q_c=\bar q-\kappa\bar\rho_c$ rebating overlap-creating seats by their
connectivity value $\rho_{ic}=\|\Pi L^+ b_{ic}\|^2$ (the same kernel again). A
Fisher-market equilibrium exists for the fractional relaxation (Eisenberg--Gale),
and the fixed-budget pivot recovers the fraction $1-1/K$ of the Groves subsidy ---
the student-side simplex tax across $K$ students, parallel to $1-1/m$. (The
connectivity-floor constraint $\lambda_2\ge\lambda_{\min}$ and the
fractional-to-integral rounding are flagged as unproven in the source.)

The impossibility (SG4) is the suite's own wall: in the symmetric fixed-supply
share class, with intensity cap $\bar\eta=\psi(\bar\tau)^{-1}\log\frac{1/s_{\min}-1}{m-1}$
forced by an IR share floor $s_{\min}$, full-fidelity grading $\tau=0$ is a Nash
equilibrium iff
\[
  u'(0)\ \le\ \beta\bar\eta\,\psi'(0)\,(1-\tfrac1m)\ =:\ \bar u ,
\]
and \emph{no} mechanism in the class implements $\tau=0$ when $u'(0)>\bar u$
(``$\alpha$ dominates''), the money-free analogue of
\citet[Thm.~4.5]{ganskominers2026}.

\paragraph{Truthful performance revelation (SG3): rejected by the reviewer gate.}
The suite attempted to turn the eigengrade \emph{residual} into a money-free
peer-prediction mechanism: each edge $e=(i,c)$ is graded by $\ge2$ graders
observing $s_e^{(k)}=P_{ic}+\omega_e^{(k)}$, scored against the leave-one-out
eigengrade prior $\mu^{-e}=\hat x_i^{-e}-\hat x_{n+c}^{-e}$.

\begin{quote}
\textbf{Proposition (independence of the leave-one-out residual).} $\mu^{-e}$ is a
function of $r_{-e}$ only, hence statistically independent of grader $k$'s private
noise $\omega_e^{(k)}$, and estimates the mean $a_i-d_c$, not
$P_{ic}=a_i-d_c+\varepsilon_{ic}$. Consequently a proper score against $\mu^{-e}$
alone elicits the prior mean $\mathbb E[P_{ic}]$, not the private signal.
\end{quote}

\methodflaw{SG3 was \emph{rejected by the reviewer gate} (logged
\textsc{reviewer-rejected 2026-06-04). The intended propriety theorem --- that the
unique expected-score maximizer is the precision-weighted blend
$r^\star=(\sigma^{-2}s_e^{(k)}+(v_e^{\mathrm{pr}})^{-1}\mu^{-e})/(\sigma^{-2}+(v_e^{\mathrm{pr}})^{-1})$,
so honest reporting is a best response --- is \emph{false in the verified
construction}. The reason is exactly the (correct) independence proposition above:
because the leave-one-out residual $\mu^{-e}$ is independent of the grader's own
signal, scoring against it elicits the \emph{prior mean}, and a \emph{constant}
report equal to the prior mean strictly beats the claimed ``honest'' blend
(reviewer numerics: constant scores $-1.130$ vs.\ the honest blend's $-1.348$ on
the same $5\times4$ instance), so honesty is strictly dominated. An
``all-compress'' profile likewise out-scores honesty ($-13.83$ vs.\ $-21.16$), and
the ``argmax reward $=$ argmax joint likelihood'' alignment claim is vacuous (the
plugged-in predictive variance $v_e$ holds for any report with that MSE). The
section is retained in the source \emph{only} as the correct independence
proposition plus a prior-art position; the incentive theorem is downgraded to an
open conjecture.}}

The suite then composes the two sides (SG5), proving existence of a composite Nash
equilibrium via Kakutani--Fan--Glicksberg (instructor side an exact concave
potential game, student side a concave pseudo-market) and signing the cross-margin
interaction (own-course complementarity, generic but non-universal cross-course
substitution --- $121$ of $2463$ cross-course additions reverse the sign). It
\emph{declines} to claim composite uniqueness or a composite welfare bound, both
flagged unproven, because the composite game is not a potential game in general.
Numerically, the suite reproduces the paper's running example exactly: the
graph-Laplacian eigengrade correlates with true ability at $0.996$ against
$0.812$ for raw GPA and $0.837$ for the raw-score average, and the kernel
$\partial W/\partial\tau_c$ matches its analytic form to $4.2\times10^{-10}$.

\paragraph{The whole ledger is visible.} Unlike Methods~1 and~2, Method~3 carries
its open obligations \emph{inside} the artifact: every gap above is tagged
\texttt{\textbackslash unproven} in the source \texttt{paper.tex} and collected in
its own appendix of open obligations, and the reviewer's rejection of SG3 is
recorded in the section heading itself. The raggedness is the honesty: the items
flagged unverified here are flagged by the method, not only by my triage.

\section{Reproducibility}
\label{app:reproducibility}

This appendix documents the artifacts behind the three methods of Section~6 and the steps needed to regenerate the quantitative claims. All code, the three skills, and the three method outputs are released in a single public repository.

\subsection{Repository and models}

The skills and worked-example artifacts are at \url{https://github.com/morankor/theorist-toolbox}. The repository contains the three Claude Code skills used here---\texttt{math-proof} (Method~1), \texttt{codex-math} (Method~2), and the \texttt{co-math-init}/\texttt{co-math-status} pair (Method~3)---together with the three method outputs as example artifacts: the single-pass writeup, the adversarial-pair writeup, and the full co-mathematician project directory described below.

Every Claude pass in this paper used \emph{Claude Opus 4.8}. Method~2 additionally invokes \emph{OpenAI Codex (gpt-5.5)} as an independent adversary: the \texttt{codex-math} skill calls Codex non-interactively at high reasoning effort and returns its findings as leads to be triaged, never as verdicts. Method~3 is driven by Claude Opus 4.8 throughout, with no second model; its discipline comes from the project structure and the reviewer gate, not from a second engine. Because both Claude and Codex are stochastic and are revised over time, exact regeneration of prose is not expected; the load-bearing claims are the numerical and symbolic checks below, which are deterministic given the published inputs.

\subsection{The co-mathematician project structure}

Method~3 produces a self-contained project directory whose layout follows the AI co-mathematician architecture of \citet{zheng2026}. At the top level it carries the living working paper \texttt{paper.tex}; a \texttt{goals.md} file holding the research question and the sub-goals SG0--SG5 with their approval status; an append-only \texttt{decisions.md} log; a \texttt{references/} directory containing the source paper \citep{ganskominers2026}; a \texttt{failed-explorations/} record; and per-project configuration in \texttt{co-math-config.json} and \texttt{.claude/settings.json}. Each sub-goal is worked in its own numbered \texttt{workstreams/W\{NNN\}-\{slug\}/} directory containing an incremental \texttt{report.md} and any verification scripts.

The reviewer gate lives in \texttt{.co-math/}: a \texttt{workstream-registry.json} and an \texttt{approvals/} directory with one record per workstream. Strict mode (\texttt{strict\_mode: true}) is enforced through the review policy, which requires reviewer approval before a workstream may be marked complete, requires citation validation, and blocks any theorem that lacks either a proof or an explicit \unproven{} flag. This gate is load-bearing rather than ceremonial: the approval record for SG3 (peer-prediction) is a \emph{rejection}. The reviewer established that the leave-one-out residual is independent of the grader's own signal, so it elicits the prior mean rather than the realized performance, and that a constant prior-mean report strictly beats the proposed ``honest'' report in the verified $5\times4$ instance. SG3 is therefore retained only as a partial section with open conjectures, and its negative outcome is recorded both in \texttt{approvals/W005-SG3.md} and in the open-obligations appendix.

\subsection{Numerical validation}

The quantitative claims are reproduced from the Gans--Kominers running example---5 students and 8 courses---by the self-contained script \texttt{validate\_running\_example.py} in the \texttt{W008-numerics} workstream. The script takes the student abilities $a_i$, course difficulties $d_c$, and observed cardinal transcript scores verbatim from Tables~2--3 of \citet{ganskominers2026}, builds the signed incidence matrix $B$ and the weighted Laplacian, and recomputes the headline correlations of their Table~4. It reproduces those figures exactly: the graph-Laplacian eigengrade correlates with true ability at $0.996$, against $0.812$ for letter grade-point average and $0.837$ for the raw-score average, and the eigengrade ranking matches the true-ability ranking.

The same script verifies the externality kernel that is the convergent finding of Section~6. It confirms numerically that the welfare gradient with respect to grading fidelity,
\[
\frac{\partial W}{\partial \tau_c}
= -\,\frac{2\tau_c}{(\sigma^2+\tau_c^2)^2}\sum_{i:(i,c)}\bigl\lVert \Pi\, L_W^{+}\, b_{ic}\bigr\rVert^2,
\]
agrees with its analytic form to $4.2\times10^{-10}$ and is everywhere nonpositive, that the softmax share pivot reproduces the money pivot ($0.3036$ at the calibrated $\eta^\star=0.3534$, $\tau^\star=0.982$) with Nash best response $0.982$, that the SG1 potential $\Phi$ matches to $3.3\times10^{-10}$ and is concave on the tested box (worst Hessian eigenvalue $-0.478$), and that the SG2 precision externality is real on the paper's own data: a thinning student swap lowers the algebraic connectivity $\lambda_2(L)$ from $0.378$ to $0.191$ and raises $\operatorname{tr}(L^{+})$ from $6.43$ to $9.24$. Companion scripts in the SG0--SG5 workstreams (\texttt{code\_verify\_identities.py}, \texttt{sg1\_potential\_check.py}, and others) check the remaining symbolic identities.

Reproduction requires only Python with NumPy. Running \texttt{validate\_running\_example.py} regenerates the Table~4 correlations and the kernel checks in a single pass; the per-workstream scripts can be run independently. We note for completeness that Method~1 ships \emph{no} numerical artifact---its limitation discussed in Appendix~A is precisely the absence of any such check---so the validation here covers Methods~2 and~3, with the running-example reproduction supplied by Method~3.

\section{Consolidated open obligations}
\label{app:open-obligations}

This appendix collects, in one place, every result that remains unproven, conditional, or rejected across the three methods of \S\ref{sec:methods}. It is a deliberate ledger: the point of a verification-first protocol is not that the outputs are complete, but that their incompleteness is recorded rather than hidden. Items are grouped by method and stated neutrally; each is a claim the corresponding writeup either flags itself or that the adversary or reviewer gate forced open. Cross-references are to the three example outputs in the repository (\url{https://github.com/morankor/theorist-toolbox}, to be made public, anonymized for review): the single-pass enrollment writeup (Method~1), the adversarial-pair grading writeup (Method~2), and the multi-agent project's \texttt{paper.tex} (Method~3).

\subsection{Method 1 (single pass): the student-side enrollment Groves mechanism}

\begin{description}[leftmargin=1.4em,style=nextline]
\item[M1.1 (no numerical verification).] Not one quantity in the writeup is checked against a computation. The marginal-informativeness identity $\partial\Psi/\partial z_e=-\|KL^{+}b_e\|^2$, the convexity of $\Psi$, the Pigouvian characterization, and the welfare-dominance claim are all derived analytically and never exercised on an instance. This is the method's defining limitation.
\item[M1.2 (individual rationality holds only for the course-contrast target).] The Clarke-pivot IR theorem (Thm.~4 of the writeup) is proved via the participation-monotonicity Lemma~1(c), which holds for the fixed-dimensional course-contrast target \textbf{(T-course)}. For the student-ability target \textbf{(T-stud)}---the eigengrade object $a-\bar a\mathbf 1_n$---adding a student introduces her own variance term, the monotonicity can fail, and IR is left conditional on an unverified ``marginal contribution $\ge 0$'' requirement. The leading specification is silently switched to \textbf{(T-course)} for the clean statements.
\item[M1.3 (budget deficit not quantified).] The mechanism is shown to run a Green--Laffont deficit; the size of that deficit, and any second-best balanced alternative, are not addressed.
\item[M1.4 (combinatorial decision relaxed, rounding not analyzed).] First-best curriculum selection is combinatorial; the convex (A-optimal-design) relaxation is offered, but the integrality gap of the rounded curriculum is asserted (``usual design guarantees'') rather than bounded.
\item[M1.5 (collusion and false-name manipulation).] Sybil enrollments to harvest the informational subsidy, and coalitional deviations, are noted as live VCG vulnerabilities and explicitly left to a future false-name-proof refinement.
\end{description}

\subsection{Method 2 (adversarial pair): the instructor-side Pigouvian transfer}

The three items M2.1--M2.3 below are claims the author first wrote and the Codex adversary then \emph{killed}; they are retained here because the corrections are part of the record, not despite being errors.

\begin{description}[leftmargin=1.4em,style=nextline]
\item[M2.1 (ex-post incentive neutrality: disproved).] The leave-one-out (``pivotal'') eigengrade was first claimed neutral in every realized sample. The adversary's $n=3$ counterexample showed realized other-student shocks make any single transcript sample-dependent; the surviving claim is first-moment / in-expectation neutrality only (Prop.~3.1 of the writeup).
\item[M2.2 (exact budget balance: proved impossible).] Balance $\sum_c t_c=0$ was claimed, then shown impossible for $m\ge 2,\ \beta>0,\ \eta>0$ via the $m$-fold mixed derivative $\partial_{1\cdots m}\big(\beta\sum_c\log s_c\big)=\beta(-1)^m m!\,\eta^m/m^m\ne 0$. AGV is named as the fallback that restores balance only in expectation, under Bayes--Nash rather than dominant strategies.
\item[M2.3 (role of the leave-one-out estimator: downgraded).] The premise that leave-out is needed for first-moment exactness was wrong: the standard estimator is already first-moment neutral. Leave-out's genuine contribution is self-influence-freeness ($\partial\hat d_c/\partial\varepsilon_{ic}=0$), not first-moment correction.
\item[M2.4 (action game, not a direct-revelation VCG).] The mechanism of \S4 is a Groves/Pigouvian corrective transfer over instructors' grade \emph{actions}. Calling it ``VCG/Clarke'' is justified only once it is embedded in a direct-revelation environment with instructors reporting cost types $\theta_c=(\alpha_c,\xi_c,\bar G_{0,c})$ and $h_c$ the Clarke pivot; that embedding is described but not constructed.
\item[M2.5 (AGV transfer not constructed).] The expected-externality remedy for M2.2 is invoked but not written down; it requires a stated prior over instructor cost types, which is not specified.
\item[M2.6 (second-moment student channel open).] Under a Bayesian (shrinkage) market, course choice still moves the \emph{precision} of the report through the Laplacian leverage, so a posterior mean can depend on the chosen course. Closing this requires a report-only or precision-equalized market; it is the precise residual the source paper hedges with ``approximate.''
\item[M2.7 (individual rationality on the unbounded grade domain).] On $\bar G_c\in\mathbb R$ the transfer becomes a subsidy for $\bar G_c<\bar G_c^\star$ and cannot be made everywhere nonpositive; ex-post IR is recovered only on a bounded grade interval with a finite entry fee, which is asserted, not derived.
\end{description}

\subsection{Method 3 (multi-agent project): the money-free report-space suite}

\begin{description}[leftmargin=1.4em,style=nextline]
\item[M3.1 (SG0 Clarke-pivot IR; existence under coercivity).] Individual rationality of the Clarke-pivot version of the money benchmark is flagged unproven; well-posedness of the efficient rule relies on Assumption~A-coerc (compact action box $\tau_c\in[0,\bar\tau]$ or coercive $u_c$), since the precision objective alone need not be coercive.
\item[M3.2 (SG1-a existence is conditional on coercivity).] The share game is an exact concave potential game, so global best response and uniqueness hold unconditionally; \emph{existence} of an equilibrium does not. A reviewer counterexample ($m=2$, linear $u_c,\psi_c$) makes $\Phi$ non-coercive and the share game has no Nash equilibrium on $[0,\infty)^m$. Existence is recovered only under the box (A-coerc) or coercivity (H-coerc) hypothesis; the unconditional existence claim is withdrawn, and the downstream composition (SG5) must invoke A-coerc/H-coerc.
\item[M3.3 (SG1 asymmetric efficiency matching).] A single intensity $\eta$ matches the $m$ social first-order conditions only in the course-symmetric environment. In the asymmetric case the mechanism pins a unique equilibrium (given coercivity) but generically not the efficient one; efficiency requires enriching the penalty profile $\psi_c$, which is described but not the exact construction proved.
\item[M3.4 (SG2 integral equilibrium and floor-preserving rounding).] Existence is proved only for the fractional Eisenberg--Gale relaxation \emph{without} the connectivity floor. An integral pseudo-market equilibrium meeting the floor $\lambda_2(L)\ge\lambda_{\min}$, and a rounding that preserves the floor, are not established; the floor is treated as a post-hoc feasibility filter.
\item[M3.5 (SG2 exact fixed-budget pivot).] The tax $\rho_{\mathrm{tax}}=1-1/K$ is established as a first-order, symmetric-environment adding-up identity only. Its exact form in an asymmetric pseudo-market, and the claim that $K/(K-1)$ is the precise (not merely first-order) inflation needed, are open---parallel to M3.3.
\item[M3.6 (SG2 efficiency-loss terms not bounded).] In the loss decomposition $(1-\rho_{\mathrm{tax}})\Delta_W+\epsilon_{\mathrm{round}}+\epsilon_{\mathrm{myopia}}$, the rounding gap $\epsilon_{\mathrm{round}}$ and price-taking myopia gap $\epsilon_{\mathrm{myopia}}$ are named but not bounded in closed form. The verification instance shows $\mathrm{tr}(\Pi L^{+}\Pi)$ and $\lambda_2(L)$ are not co-monotone, so even the no-tax, no-myopia fractional optimum of a $\lambda_2$- or $\rho$-priced market need not coincide with the first-best; full money-free first-best implementation is therefore not achieved.
\item[M3.7 (SG3 rejected by the reviewer gate).] The peer-prediction route to truthful performance revelation was rejected. The retained content is the correct independence proposition (Prop.~\ref{app:m3}): the leave-one-out residual $\mu^{-e}$ is independent of the grader's own signal and estimates the prior mean $a_i-d_c$, so a proper score against it elicits the prior mean and a constant report strictly beats honesty (constant-prior-mean report scores $-1.130$ vs.\ the claimed honest blend $-1.348$). The propriety/honesty theorem, the likelihood/Groves/welfare alignment, exclusion of uninformative and colluding equilibria, the minimal multi-grader coverage requirement, and composition with SG1 all remain open.
\item[M3.8 (SG4 global sufficiency).] The ``$\alpha$ dominates'' wall is fully established in its necessity and impossibility directions (i),(iii). The sufficiency direction (ii) rests on an unproven global-best-response step: that the first-order corner condition implies a global best response over $[0,\bar\tau]$, which holds under a monotone-ratio condition $u'/\psi'$ nonincreasing that is stated but not established for the relevant $(u,\psi)$ class.
\item[M3.9 (SG4 mechanism-independence).] The impossibility is proved for the fixed-supply \emph{softmax}-share class with a common penalty $\psi$ and a single intensity $\eta$. Whether every fixed-supply, IR, share-feasible no-money mechanism inherits a finite corner-pivot cap---the genuinely mechanism-independent, Gibbard--Satterthwaite-type statement---is open, pending a revelation/taxation-principle argument. The robustness of the cap $\bar\eta$ to coordinated multi-instructor (rather than single-deviator) compression is stated to weakly enlarge the impossibility region but not formally bounded.
\item[M3.10 (SG5 uniqueness).] The composite two-sided game has a pure-strategy Nash equilibrium (Kakutani--Fan--Glicksberg), but the equilibrium is not shown unique: the coupling $x\mapsto\eta(x)$ is not a contraction and no joint potential exists, so uniqueness would require an unestablished diagonal-strict-concavity / Rosen monotone-game condition.
\item[M3.11 (SG5 welfare).] No welfare bound is proved for the composite equilibrium relative to the money-VCG benchmark: SG1 matches the money pivot only in the symmetric regime at the calibrated $\eta^\star$, and under endogenous asymmetric enrollment the realized $\eta(x^\dagger)$ need not match every course's pivot, leaving composite informativeness loss unbounded.
\item[M3.12 (SG5 own-course complementarity not a closed inequality).] Own-course complementarity (more enrollment strengthens an instructor's fidelity incentive) is verified numerically on $297/297$ instances but not proved: appended-term dominance over the retained-term rank-one update is not established as an inequality. The cross-course sign is environment-dependent ($95.1\%$ substitution; $121$ of $2463$ additions are complements), so no global cross-margin monotonicity is available.
\end{description}

\noindent Two features of this ledger bear on the methodological argument of \S\ref{sec:findings}. First, Method~1---the most polished output---contributes the entry M1.1 that none of its results were ever computed, while Method~3, whose draft carries its open items inline as \texttt{\textbackslash unproven} flags, supplies the longest and most specific list precisely because its reviewer gate forced the disclosures. Second, the items the adversary or gate killed (M2.1--M2.3, M3.2, M3.7) are not failures of the protocol but its intended product: each is a false claim that a single unverified pass would have shipped.

\end{document}